\documentclass[sigconf]{acmart}
\usepackage[normalem]{ulem}
\usepackage{amsmath}
\usepackage{multirow}
\usepackage{epstopdf}
\usepackage{graphicx}
\usepackage{balance}
\usepackage{makecell}

\usepackage[normalem]{ulem}
\useunder{\uline}{\ul}{}
\usepackage{enumitem}
\useunder{\uline}{\ul}{}

\AtBeginDocument{%
  \providecommand\BibTeX{{%
    \normalfont B\kern-0.5em{\scshape i\kern-0.25em b}\kern-0.8em\TeX}}}

\copyrightyear{2023}
\acmYear{2023}
\setcopyright{acmlicensed}\acmConference[KDD '23]{Proceedings of the 29th ACM SIGKDD Conference on Knowledge Discovery and Data Mining}{August 6--10, 2023}{Long Beach, CA, USA}
\acmBooktitle{Proceedings of the 29th ACM SIGKDD Conference on Knowledge Discovery and Data Mining (KDD '23), August 6--10, 2023, Long Beach, CA, USA}
\acmPrice{15.00}
\acmDOI{10.1145/3580305.3599884}
\acmISBN{979-8-4007-0103-0/23/08}

\usepackage{amsmath}
\usepackage{graphicx} 
\usepackage{float} 
\usepackage{subfigure} 
\usepackage{setspace}

\begin{document}

\title{PEPNet: Parameter and Embedding Personalized Network for Infusing with Personalized Prior Information}

\author{Jianxin Chang}
\affiliation{
  \institution{Kuaishou Technology}
  \city{Beijing}
  \country{China}}
\email{changjianxin@kuaishou.com}

\author{Chenbin Zhang}
\authornote{Equal contribution. Author ordering determined by coin flip.}
\affiliation{
  \institution{Kuaishou Technology}
  \city{Beijing}
  \country{China}}
\email{zhangchenbin@kuaishou.com}

\author{Yiqun Hui}
\affiliation{
  \institution{Kuaishou Technology}
  \city{Beijing}
  \country{China}}
\email{huiyiqun@kuaishou.com}

\author{Dewei Leng}
\affiliation{
  \institution{Kuaishou Technology}
  \city{Beijing}
  \country{China}}
\email{lengdewei@kuaishou.com}

\author{Yanan Niu}
\affiliation{
  \institution{Kuaishou Technology}
  \city{Beijing}
  \country{China}}
\email{niuyanan@kuaishou.com}

\author{Yang Song}
\affiliation{
  \institution{Kuaishou Technology}
  \city{Beijing}
  \country{China}}
\email{yangsong@kuaishou.com}

\author{Kun Gai}
\affiliation{
  \institution{Unaffiliated}
  \city{Beijing}
  \country{China}}
\email{gai.kun@qq.com}

\renewcommand{\shortauthors}{Jianxin Chang et al.}

\begin{abstract}
With the increase of content pages and interactive buttons in online services such as online-shopping and video-watching websites, industrial-scale recommender systems face challenges in multi-domain and multi-task recommendations.
The core of multi-task and multi-domain recommendation is to accurately capture user interests in multiple scenarios given multiple user behaviors.
In this paper, we propose a plug-and-play \textit{\textbf{P}arameter and \textbf{E}mbedding \textbf{P}ersonalized \textbf{Net}work (\textbf{PEPNet})} for multi-domain and multi-task recommendation.
PEPNet takes personalized prior information as input and dynamically scales the bottom-level Embedding and top-level DNN hidden units through gate mechanisms. 
\textit{Embedding Personalized Network (EPNet)} performs personalized selection on Embedding to fuse features with different importance for different users in multiple domains. 
\textit{Parameter Personalized Network (PPNet)} executes personalized modification on DNN parameters to balance targets with different sparsity for different users in multiple tasks. 
We have made a series of special engineering optimizations combining the Kuaishou training framework and the online deployment environment.
By infusing personalized selection of Embedding and personalized modification of DNN parameters, PEPNet tailored to the interests of each individual obtains significant performance gains, with online improvements exceeding 1\% in multiple task metrics across multiple domains.
We have deployed PEPNet in Kuaishou apps, serving over 300 million users every day. 
\end{abstract}

\ccsdesc[500]{Information systems~Personalization}
\ccsdesc[500]{Information systems~Recommender systems}
\ccsdesc[500]{Computing methodologies~Neural networks}

\keywords{Multi-Domain Learning; Multi-Task Learning; Personalization; Recommender System}
\maketitle

\vspace{-4px}
\section{Introduction}
\begin{figure}[t]
  \centering
\includegraphics[width=0.47\textwidth]{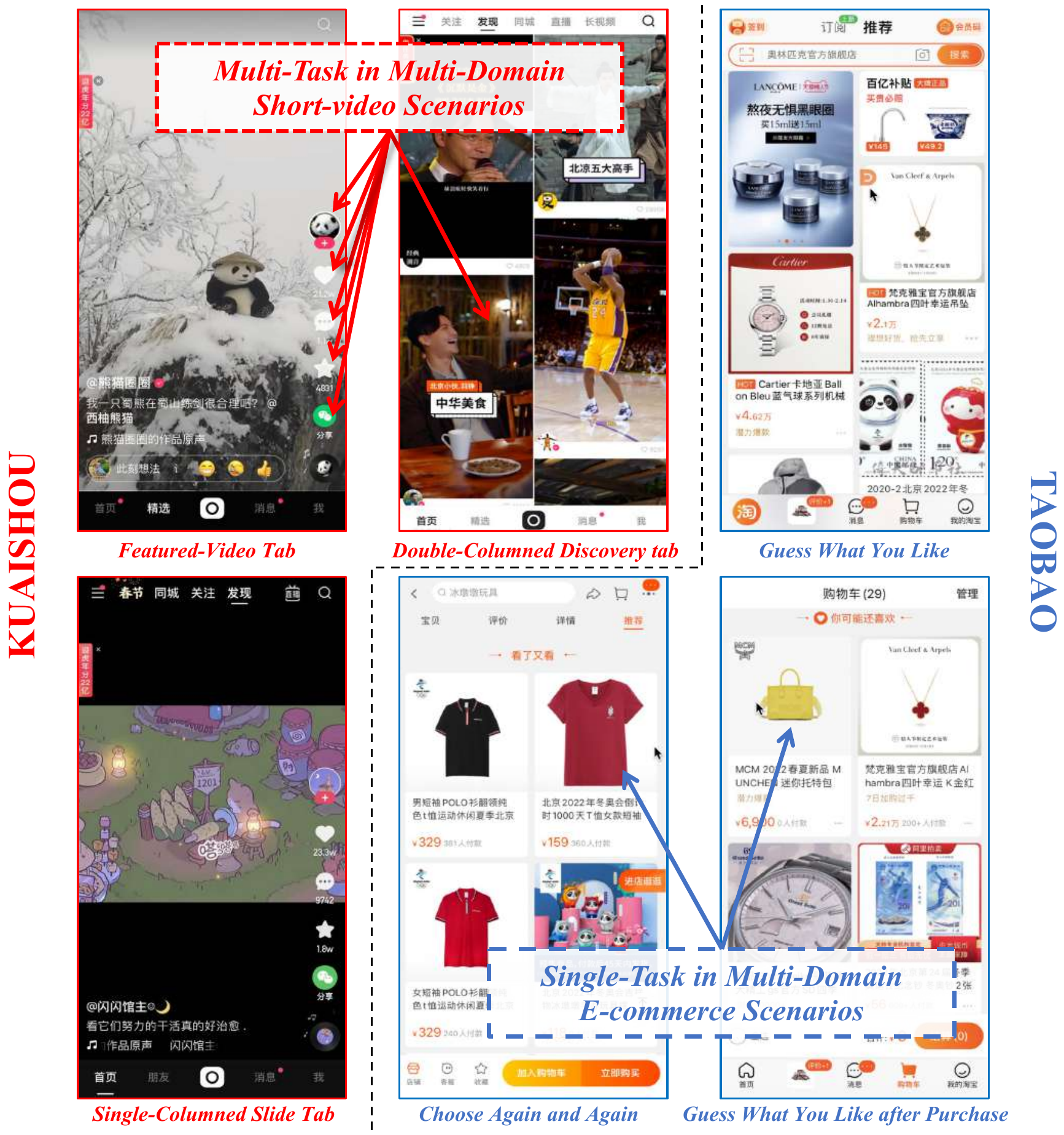}
  \caption{Comparison of short-video scenarios in Kuaishou and e-commerce scenarios in Taobao. Both of them make recommendation for different domains. In addition, multiple tasks are carried out for each domain in Kuaishou, e.g., {\em like, follow, forward, collect, and comment} for short videos.}
  \vspace{-10px}
  \label{fig:group}
\end{figure}

Traditional recommendation models focus on the single prediction task (e.g. CTR) in a single domain\cite{cheng2016wide, lian2018xdeepfm, wang2021dcn}, which is training using examples collected from a \textbf{single domain} and serving the prediction of a \textbf{single task}. 
However, in real-world applications, the need for recommendation are fragmented across different scenarios.
As the number of content pages increases, recommender systems face the critical problem that data fragments are located in \textbf{multiple domains}.
For example, Taobao\footnote{\url{https://www.taobao.com/}} has scenarios such as \textit{pre-purchase(Guess What You Like)}, \textit{in-purchase(Choose Again and Again)}, and \textit{post-purchase(Guess What You Like after Purchase)}, as shown in Figure~\ref{fig:group}. 
And Kuaishou\footnote{\url{https://www.kuaishou.com/}} has scenarios such as \textit{Featured-Video Tab}, \textit{Double-Columned Discovery Tab}, and \textit{Single-Columned Slide Tab}.
In addition, multiple buttons are usually designed on each page for users to interact with.
To leverage user feedback and provide a better experience, recommender systems need to capture various behavior preferences of users, modeling the probability of user interactions with different targets in \textbf{multiple tasks}. 
For example, Kuaishou provides users with various interaction targets in Figure~\ref{fig:group}, such as \textit{like}, \textit{follow}, \textit{forward}, \textit{collect}, and \textit{comment}.

Since there are overlapping users and items in different scenarios, the multiple domains have commonalities.
And different targets are functionally related, so there are dependencies between multiple tasks.
Training separate models for each task in each domain is not only unacceptable in terms of deployment cost and iterative efficiency, but also not utilizing the full amount of data and ignoring the commonalities between the data can lead to suboptimal performance.
However, mixing all the data directly and training with a unified model ignores the differences between domains and tasks.
The inability to align and fuse features with different semantics and importances will result in \textbf{domain seesaw}~\cite{sheng2021one} due to the various distributions of user behaviors and item candidates in multiple scenarios.
Since different targets have distinctive sparsity and influence each other, the inability to balance interdependent targets in multiple tasks can lead to the \textbf{task seesaw}~\cite{tang2020progressive}.

\begin{figure}[t]
  \centering
  \hspace{-0.4cm}
\includegraphics[width=0.43\textwidth]{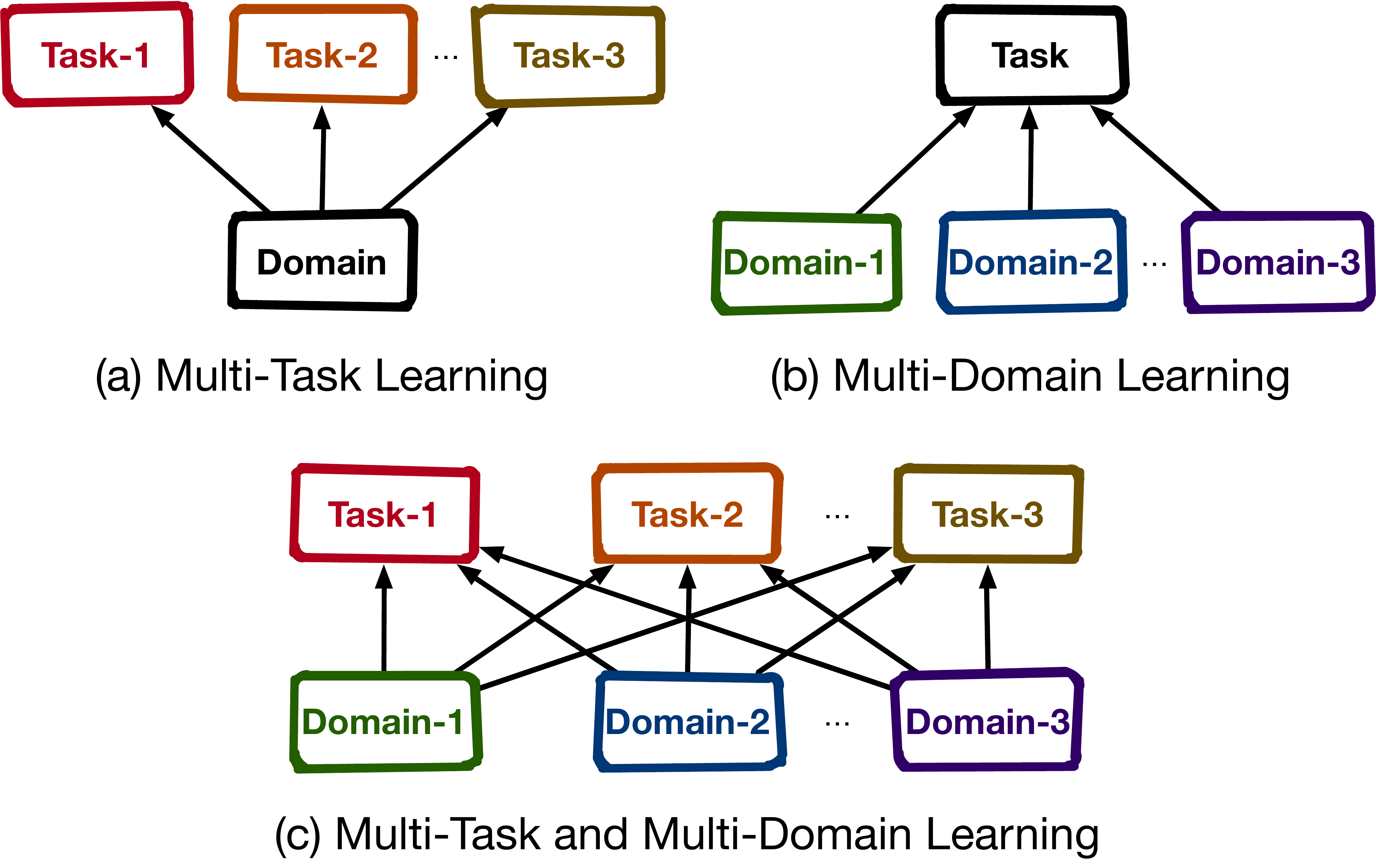}
  \caption{Compared with multi-task learning or multi-domain learning, multi-task and multi-domain learning is more important in real applications and is more complex.}
  \vspace{-12px}
  \label{fig:mtl_mdl}
\end{figure}

At present, \textbf{multi-domain learning} and \textbf{multi-task learning} have made great progress in recommender systems~\cite{zhang2016deep, qu2016product, cheng2016wide, wang2017deep, wang2021dcn}. 
But in real applications, we cannot simply and directly reuse multi-domain or multi-task learning methods in multi-domain and multi-task joint settings, respectively.
Multi-domain methods focus on aligning the feature semantics under different domains, but ignore target dependencies in the label space under multi-task settings.
Multi-task methods focus on fitting target distributions of different tasks, but ignore the semantic differences in the feature space under multi-domain settings.
As shown in Figure~\ref{fig:mtl_mdl}, compared with separate multi-task learning or multi-domain learning, multi-task learning and multi-domain learning occur simultaneously in real applications and are more complex.
On the one hand, there are gap in the feature semantics and importance between different domains of the same task and the same domain of different tasks. 
On the other hand, different tasks within the same domain and the same task within different domains have various target sparsity and interdependence.
Different from task seesaw phenomenon and domain seesaw phenomenon, we call it \textbf{the imperfectly double seesaw phenomenon}. 
The phenomenon is more severe in industry-scale recommender systems as the number of domains and tasks increases.
Due to the requirement for high efficiency and low cost in real industries, a plug-and-play network is urgently needed to solve the challenges of multi-domain and multi-task.
Personalization modeling is the core of recommender systems.
Augmenting personalization of the model helps capture the degree of user preference for items in different situations.
Multi-domain and multi-task settings can be viewed as users interacting with items in different situations, so more accurate personalization estimates can alleviate the imperfectly double seesaw problem.
But simply using personalized priors information as the bottom input, the effect becomes extremely weak after their signal passes through deep networks to the top layer.
How to infuse personalized priors into the model in the right place and in the right way is critical and worth exploring, especially for multiple domains and tasks.

To address this issue, we propose a \textit{\textbf{P}arameter and \textbf{E}mbedding \textbf{P}ersonalized \textbf{Net}work (\textbf{PEPNet})} for the multi-task and multi-domain recommendation, 
which fully exploits the relationship between tasks and eliminates domain bias via augmenting personalization. 
Compared with existing works in multi-task learning \cite{ma2018modeling, tang2020progressive} and multi-domain learning \cite{sheng2021one, li2020ddtcdr}, PEPNet is an efficient plug-and-play network. 
PEPNet takes features with personalized prior information as input and dynamically scales the bottom-layer Embedding and the top-layer DNN hidden units in the model through the gate mechanism, which are called domain-specific EPNet and task-specific PPNet. 
\textit{Embedding Personalized Network (EPNet)} adds domain-specific personalized information to the bottom layer to generate personalized Embedding gates. 
And the Embedding gates are used to perform personalized selection on the original Embedding from multiple domains to get the personalized Embedding.
\textit{Parameter Personalized Network (PPNet)} concatenates personalized information about the user and item with the input of DNN in each task tower to obtain the personalized gate scores. 
Then element-wise product is executed with DNN hidden units to make personalized modifications to DNN parameters.
By mapping personalized priors to scaling weights ranging from 0 to 2, EPNet selects Embedding to fuse features with different importance for different users in multiple domains, and PPNet modifies DNN parameters to balance targets with different sparsity for different users in multiple tasks. 

The contributions of this work can be summarized as follows:
\begin{itemize}[leftmargin=*,partopsep=0pt,topsep=0pt]
\setlength{\itemsep}{0pt}
\setlength{\parsep}{0pt}
\setlength{\parskip}{0pt}
  \item We propose a Parameter and Embedding Personalized Network (PEPNet) tailored to the interests of each individual. PEPNet is an efficient, low-cost deployment and plug-and-play method that can be injected into any model. We evaluate PEPNet and other SOTA methods on the industrial short-video dataset, and extensive experiments demonstrate the effectiveness of our method in mitigating the imperfectly double seesaw phenomenon.
  \item We deploy PEPNet in the recommendation system of Kuaishou, serving more than 300 million daily active users (DAU). The deployment of PEPNet brings a more than 1\% increase in watch time and around 2\% improvement on multiple interactive targets. Our method can be generalized to other setups, and researchers 
  can benefit from the lessons learned in our deployment. 
\end{itemize}

\begin{figure*}[t]
  \centering
  \vspace{-0.3cm} 
  \includegraphics[width=0.82\textwidth]{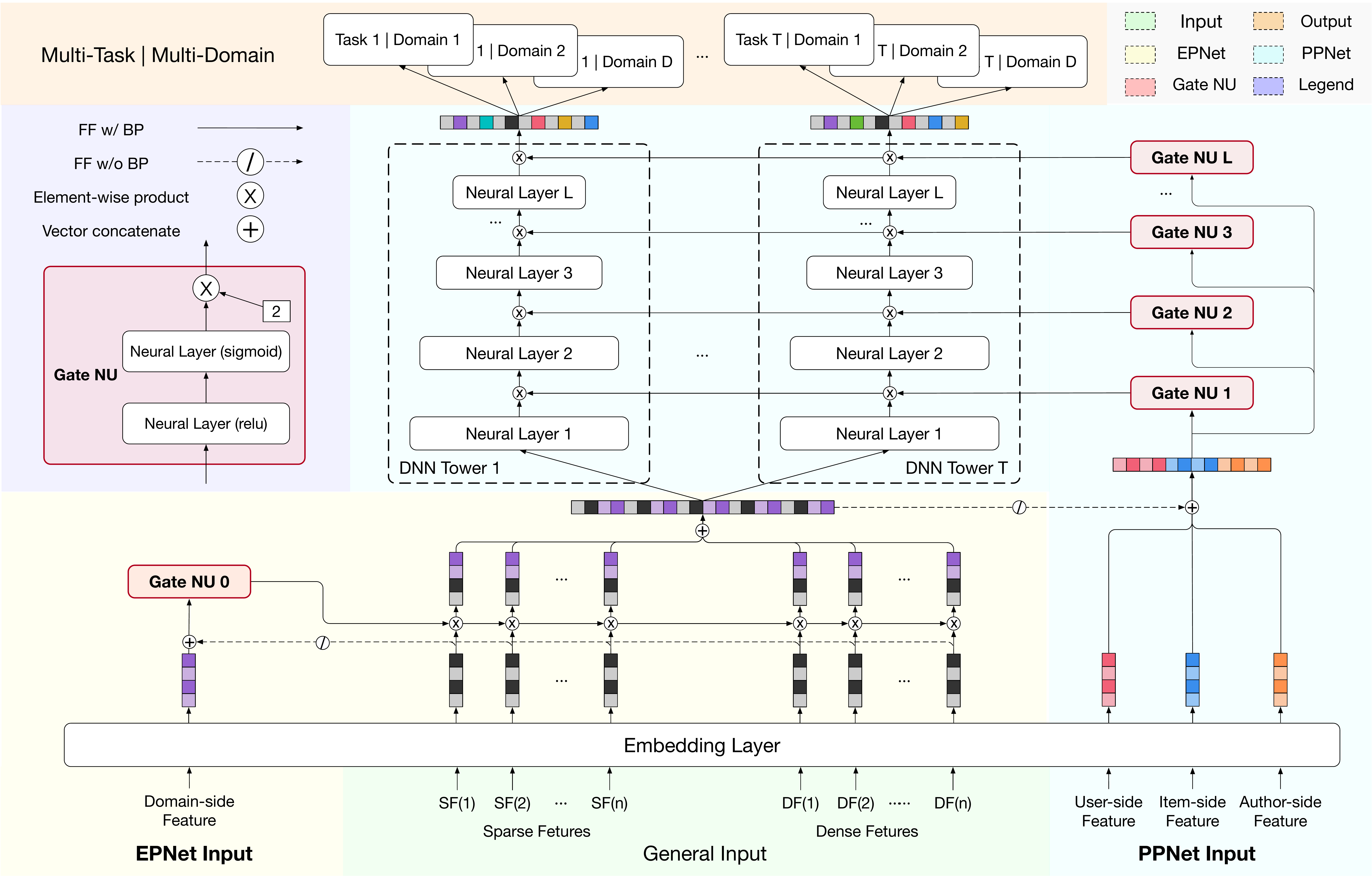}
  \vspace{-0.2cm} 
  \caption{
PEPNet consists of Gate NU, EPNet and PPNet. 
Gate NU is the basic unit that utilizes prior information to generate personalized gates and adaptively amplifies valid signals. 
EPNet performs personalized selection on Embedding to fuse features with different importance for different users in multiple domains. 
PPNet executes personalized modification on DNN parameters to balance targets with different sparsity for different users in multiple tasks. 
The same set of multi-targets is estimated in multiple domains. 
PEPNet, with few parameters and fast convergence speed, can be plugged and played into any network.
}
  \vspace{-0.3cm} 
  \label{fig:PEPNet}
\end{figure*}

\section{Methodology} \label{sec:method}
This section presents the detailed design for alleviating the imperfectly double seesaw problem. 
We elaborate on problem formulation, network structure of the proposed PEPNet and deployment in Kuaishou, one of the largest short-video platforms in China.

\subsection{Problem Formulation}
Here we define the notations and problem settings of our study. The model uses sparse/dense inputs such as user historical behavior, user profile features, item features, context features and so on. The predicted target ${\hat{y}}_{t}$ is the user $u$ preference score on an item $i$ for the $t$-th task in domain $d$, which is calculated via: 

\begin{equation}
\begin{aligned}
{\hat{y}}_{t}=F( \{E\left(u_{1}\right), \ldots, E\left(u_{m}\right)  \oplus E\left(i_{1}\right), \ldots,   \\
E\left(i_{n}\right)  \oplus
E\left(c_{1}\right), \ldots, E\left(c_{o}\right)\}_{d} )
\end{aligned}
\end{equation}
where ${u_{1},...,u_{m}}$ indicate user features including the user historical behavior, user profile, and user ID, etc. ${i_{1},...,i_{n}}$ indicate the item features including item category, item ID (iid), and author ID (aid), etc. ${c_{1},...,c_{o}}$ indicate the other features which include the context feature and combine feature. 
$m$, $n$ and $o$ refer to the number of user features, item features and other features, respectively.
$E(*)$ means the sparse/dense features are mapped to the learnable embedding by the embedding layer after the bucketing algorithm, and $\oplus$ indicates concatenation.
$\{\}_{d}$ represents the examples from domain $d$. 
${\hat{y}}_{t}$ denotes the output score for task $t$.
$F$ is the recommendation model, which is essentially a learnable prediction function.

In the real world, the item candidate pool and part of users are shared in multiple scenarios. 
However, due to different consumption purposes, user behavior tendencies towards the same item will change in different scenarios.
To better capture user tendency to multiple behaviors and their connection in multiple scenarios, 
the recommender $f$ needs to make predictions for multi-task $T$ in multiple domains $D$ simultaneously. 
So the multi-domain and multi-task recommendation problem can be formulated as: $x_d \rightarrow {\hat{y}}_{t}$, where $x_d$ is the feature of examples collected from each domain $d \in D$, and ${\hat{y}}_{t}$ is the prediction score for each task $t \in T$.

\subsection{Network Structure}

Figure ~\ref{fig:PEPNet} illustrates the network structure of our proposed PEPNet model.
The overall architecture is made up of the following three parts, which we will elaborate on one by one. 

\begin{itemize}[leftmargin=*,partopsep=0pt,topsep=0pt]
    \setlength{\itemsep}{0pt}
    \setlength{\parsep}{0pt}
    \setlength{\parskip}{0pt}
\item \textbf{Gate Neural Unit.} 
Gate NU, the basic unit of EPNet and PPNet, is a gating structure that handles more prior information with different personalized semantics for injection into the model.
\item \textbf{Embedding Personalized Network.} 
EPNet takes personalized domain-specific information as input of Gate NU and performs personalized selection on Embedding to fuse features with different importance for different users in multiple domains. 
\item \textbf{Parameter Personalized Network.} 
PPNet uses personalized information about the user/item to generate gates and executes personalized modification on DNN parameters to balance targets with different sparsity for different users in multiple tasks. 
\end{itemize}

\subsubsection{\textbf{Gate Neural Unit(Gate NU)}}
Inspired by the LHUC algorithm proposed in the field of speech recognition~\cite{swietojanski2016learning}, PEPNet introduces a gating mechanism called Gate Neural Unit, which allows personalized prior information to be injected into the network.
LHUC, which focuses on learning speaker-specific hidden unit contributions, improves the accuracy of speech recognition for different speakers by scaling the model's hidden layers with personalized contributions.
However, LHUC essentially uses user ID as the personalized identifiers, ignoring other abundant personalized prior information, such as user's age, gender, and other profiles.
In addition, in recommendation systems that match users and items, item information is also crucial, such as item's ID, category, and author.
Numerous studies ~\cite{zhou2018din, zhou2019dien, PiZZWRFZG2020SIM} have shown that users express different personalized preference patterns for different items.

Therefore, we propose the Gate Neural Unit, short for Gate NU, to handle more prior information with different personalized semantics and inject it into the model.
Gate NU, also referred to as $\mho$ later, consists of two neural network layers.
We denote the inputs of Gate NU as $\mathbf{x}$, and formulate the first layer as follows, 
\begin{equation}
\mathbf{x'}=Relu\left(\mathbf{x} \mathbf{W}+\boldsymbol{b}\right),
\label{equ:2}
\end{equation}
where $\mathbf{W}$ and $\boldsymbol{b}$ are learnable weight and bias. $Relu$ is choosed as the non-linear activation function.
The first layer is used to cross features with various prior information. Then, we customize the generation of gate scores through the second layer as follows,
\begin{equation}
\boldsymbol{\delta} =\gamma * Sigmoid\left(\mathbf{x'} \mathbf{W'}+\boldsymbol{b'}\right), \boldsymbol{\delta} \in [0, \gamma].
\label{equ:3}
\end{equation} 
The output $\mathbf{x'}$ of the first layer is fed into the input of the second layer. $\mathbf{W'}$ and $\boldsymbol{b'}$ are the trainable weight and bias in the second layer.
$Sigmoid$ function is used to generate gate vectors $\boldsymbol{\delta}$, which limits the output to $[0, \gamma]$. $\gamma$ is the scaling factor that is set as 2. 
 
From Equation~\ref{equ:2} and ~\ref{equ:3}, Gate NU uses the prior information $\mathbf{x}$ to generate personalized gates $\boldsymbol{\delta}$, adaptively controlling the importance of prior information, and uses the hyperparameter $\gamma$ to further squash and double the effective signal. 
Next, we elaborate on how to use Gate NU in EPNet and PPNet to selectively inject important prior information into crucial positions of the model.

\subsubsection{\textbf{Embedding Personalized Network(EPNet)}}
In industrial-scale recommendation systems, the embedding tables are huge, especially for ID features. 
To save computation and memory costs, the share-bottom embedding structure is widely used as follows,
\begin{equation}
    \mathbf{E} = E(\mathcal{F}_S) \oplus E(\mathcal{F}_D),
\end{equation}
where $\mathcal{F}_S$ are sparse features and $\mathcal{F}_D$ are dense features.
As the general input, they are transformed into the learnable embedding $\mathbf{E}$ through the embedding layer $E(*)$.

Due to sharing the embedding layer for training samples from various domains, there are several drawbacks in practice, as it emphasizes commonalities while neglecting differences between multiple domains.
EPNet, on the basis of the shared embedding layer, injects domain-specific personalized prior information into the embedding with low cost, i.e. few parameters and fast convergence speed.
We use domain-side features $E(\mathcal{F}_d) \in \mathbb{R}^{k}$ as the input of EPNet, 
including domain ID and domain-specific personalized statistical features, such as the count of user behaviors and item exposures in each domain. 
$\mho_{ep}$ is the Gate NU of EPNet in the embedding layer, and its output $\mathbf{\delta}_{domain} \in \mathbb{R}^{e} $ is given by
\begin{equation}
    \boldsymbol{\delta}_{domain} = \mho_{ep}(E(\mathcal{F}_d)
    \oplus (\oslash (\mathbf{E}))),
\end{equation}
where we concatenate general embedding $\mathbf{E} \in \mathbb{R}^{e}$ with the input, 
but without using gradient backpropagation, denoted as $\oslash(*)$.
Next, we use the external Gate NU to perform the personalized transformation on embedding $\mathbf{E}$ without changing the original embedding layer, aligning features with different importance for different users in multiple domains. The transformed embedding is,
\begin{equation}
    \mathbf{O}_{ep} = \boldsymbol{\delta}_{domain} \otimes \mathbf{E},
\end{equation}
where $\mathbf{O}_{ep} \in \mathbb{R} ^ {e}$, and $\otimes$ denotes the element-wise product.
Note that when there are many input features and large vector dimensions in the embedding layer, the vector-wise product is optional.

\subsubsection{\textbf{Parameter Personalized Network(PPNet)}}
Existing multi-task recommenders~\cite{ma2018modeling, tang2020progressive} focus on using complex modules to model multi-task representations. 
When fitting multi-task labels based on multi-task representations, they all use DNN Towers, that is, stacked neural network layers.
However, the parameters of DNN Towers are shared by all users.
Due to the inconsistency of different users' preferences for various behaviors, the lack of personalized parameters will make it hard for the model to balance multiple tasks, inevitably leading to the performance seesaw.

To address this issue, we propose PPNet to modify DNN parameters in multi-task learning, building a DNN model tailored to each user's interests.
We use user/item/author-side features ($\mathcal{F}_u$/$\mathcal{F}_i$/$\mathcal{F}_a$) as personalized priors for PPNet, such as user ID, item ID, author ID (the producer of short videos in Kuaishou), and other side information features, e.g., user age/gender, item category/popularity, etc. Specifically, the detailed structure of PPNet is as follows:
\begin{equation}
   \begin{aligned}
    \mathbf{O}_{prior} &= E(\mathcal{F}_u) \oplus E(\mathcal{F}_i) \oplus E(\mathcal{F}_a), \\
    \boldsymbol{\delta}_{task} & = \mho_{pp}(\mathbf{O}_{prior} \oplus (\oslash (\mathbf{O}_{ep}))).
 \end{aligned} 
\end{equation}
We concatenate the output of EPNet $\mathbf{O}_{ep}$ with the personalized prior $\mathbf{O}_{prior}$ as the input of $\mho_{pp}$, which is Gate NU in PPNet.
To avoid affecting the embedding updated in EPNet, we perform the operation of stop gradient $\oslash$ on $\mathbf{O}_{ep}$. 
Next, we use the element-wise product based on Gate NU output $\boldsymbol{\delta}_{task}$ to double and squash the hidden contributions $\mathbf{H}$ in each layer of the DNN as follows:
\begin{equation}
    \mathbf{O}_{pp} = \boldsymbol{\delta}_{task} \otimes \mathbf{H},
\end{equation}
where $\mathbf{H} = [H_1, \ldots, H_T]$. In each DNN layer, $H_t \in \mathbb{R}^{h}$ denotes the hidden unit of the $t$-th task tower.
Note that $\boldsymbol{\delta}_{task} \in \mathbb{R}^{h*T}$ is applied to the hidden layer units of $T$ tasks after being split into $T$ vectors with dimension $h$.
Similarly, $\mathbf{O}_{pp}$ after splitting represents the $h$-dimensional PPNet outputs in T tasks.

Furthermore, we integrate PPNet into all DNN layers to fully personalize DNN parameters, balancing targets with different sparsity for different users in multiple tasks, formulated as follows,
\begin{equation}
 \begin{aligned}
   \mathbf{O}_{pp}^{(l)} &= \boldsymbol{\delta}_{task}^{(l)} \otimes \mathbf{H}^{(l)}, \\
   \mathbf{H}^{(l+1)} &= f(\mathbf{O}_{pp}^{(l)} \mathbf{W}^{(l)}+\boldsymbol{b}^{(l)}), l \in \{1,...,L\},
 \end{aligned}
\end{equation}
where $L$ is the number of DNN layers of task towers and $f$ is the activation function. For the first $L-1$ layers, the activation function $f$ uses Relu. $f$ in the last layer is Sigmoid without amplification coefficients $\gamma$, which is different from Gate NU. 
After obtaining prediction scores for multiple tasks on multiple domains in the last layer, the binary cross-entropy is employed for optimization.

\subsection{Engineering Optimization Strategy}
To deploy PEPNet in Kuaishou's large-scale recommendation scenarios, we make the following engineering optimization strategies:

\begin{itemize}[leftmargin=*,partopsep=0pt,topsep=0pt]
\setlength{\itemsep}{0pt}
\setlength{\parsep}{0pt}
\setlength{\parskip}{0pt}
    \item \textbf{Feature elimination strategy}:
    In large-scale recommendation systems, mapping each feature to an embedding vector quickly fills up the memory resources of the server. To avoid exhausting memory in servers storing embeddings, we design a conflict-free and memory-efficient Global Shared Embedding Table (GSET). Unlike traditional cache elimination strategies such as LFU and LRU, which focus on maximizing cache hit rate, GSET adopts a feature score elimination strategy to prevent low-frequency features from entering and exiting the system, which could negatively impact system performance. By effectively managing embedding vectors, memory usage can be kept below a predetermined threshold, ensuring long-term system performance. 
    
    \item \textbf{Online synchronization strategy}: 
    We refer to the minimum unit of training in online learning as a "pass".
    In each pass, the updated parameters of the DNN are fully synchronized online.
    However, due to the large number of users and items, it is not feasible to fully synchronize the embeddings.
    Although new users and items continue to appear, older ones may expire or become cold.
    Fully synchronizing the updated embeddings in each pass would increase the redundancy of the system, bringing additional storage, computation, and communication costs.
    To address this issue, we implement two strategies to synchronize the required embeddings in each pass.
    The first strategy is to set a quantity limit for each feature to prevent excessive synchronization of embeddings for any individual feature.
    The second strategy is to set an expiration time for the embeddings, only synchronizing those that are frequently updated and not synchronizing those that have not reached the designated update frequency.
    
    \item \textbf{Offline training strategy}:
In short-video scenarios, the updating of embeddings is more frequent than DNN parameters, especially for ID features.
To better capture changes in the bottom-layer Embedding and stably update the top-layer DNN parameter in the case of online learning, we train the Embedding and the DNN parameter separately and adopt different update strategies. 
In the Embedding layer, we use the AdaGrad optimizer and the learning rate is set to 0.05.
While the DNN parameters are updated by the Adam optimizer with the learning rate 5.0e-06.
\end{itemize}

\vspace{-4px}
\section{EXPERIMENT}
In this section, we conduct extensive experiments to evaluate PEPNet, with the purpose of answering the following questions.

\begin{itemize}[leftmargin=*,partopsep=0pt,topsep=0pt]
    \setlength{\itemsep}{0pt}
    \setlength{\parsep}{0pt}
    \setlength{\parskip}{0pt}
\item \textbf{RQ1:} 
  How does the proposed method perform compared with state-of-the-art recommenders? What about the performance in multi-task and multi-domain scenarios?
\item \textbf{RQ2:} 
  Can PPNet and EPNet in the proposed method address the imperfectly double seesaw problems in multi-task and multi-domain recommendation, respectively?
\item \textbf{RQ3:} 
What is the effect of different components and implementations in the proposed method?
\item \textbf{RQ4:}
How does PEPNet perform in real online scenarios?
\end{itemize}

\vspace{-3px}
\subsection{Experimental Settings}
\subsubsection{Datasets and Metrics.} 
To evaluate PEPNet in real-world scenarios that suffer from the imperfectly double seesaw problem, we collect an industrial dataset with rich domains and tasks from Kuaishou.
We extract a subset of the logs from Sept. 11th to Sept. 22nd, 2022, a total of 12 days.
We consider three domains that are the \textbf{Double-Columned Discovery Tab}, the \textbf{Featured-Video Tab}, and the \textbf{Single-Columned Slide Tab}, annotated as Domain A, B and C in our experiments.
Six types of user interactions are predicted as binary targets in multiple tasks, namely \textbf{Like}, \textbf{Follow}, \textbf{Forward}, \textbf{Hate}, \textbf{Click} and \textbf{EffView}.
Click in single-columned tabs is defined as watching for longer than 3 seconds to simulate the click behavior that does not exist in immersive tabs. 
EffView, short for {\em effective view}, is defined as 1 if the watch time reaches 50\% percentile or more of all samples, and 0 otherwise.

We use data from the first 10 days as the training set, the 11th day for validation, and the last day for the test. 
We further filter out users who have fewer than 10 interactions and items that are interacted with by fewer than 10 users.
And we evaluate the models using two widely-adopted accuracy metrics including AUC and GAUC~\cite{zhou2018din}.
Statistics of the dataset are summarized in Table \ref{tab::dataset}, including the basic information, the sparsity of each task in each domain, and the overlap of users and exposed items across domains.
Although domains share the same item pool and contain many overlapping users, it can be observed that item exposures and user behaviors differ across different domains.
This indicates that users have different behavioral intentions in multiple domains and experience a differentiated consumption ecosystem.

\renewcommand\arraystretch{0.90}
\begin{table}
	\caption{Statistics of the dataset used in experiments include the basic information, the sparsity of each task in each domain, and the overlap of users and exposed items across domains.
    Users have different behavioral intentions in multiple domains and experience a differentiated ecosystem.
    }
	\vspace{-10px}
	\label{tab::dataset}
	\small
	\setlength{\tabcolsep}{1mm}
 {\begin{tabular}{llcccc}
    \toprule
     & Domains
     & \makecell[c]{Discovery Tab \\ (Domain A)} 
     & \makecell[c]{Featured-Video Tab \\ (Domain B)} 
     & \makecell[c]{Slide Tab \\ (Domain C)} \\
    \midrule
    \multirow{3}{*}{Basic}  
    & Users & 76k & 110k & 88k \\
    & Items & 9,474k & 5,205k & 5,588k \\
    & Instances & 48,037k & 68,348k & 78,197k \\
    \midrule
    \multirow{7}{*}{\makecell[l]{Task \\ Sparsity}} 
    & Like & $3.68\%$ & $2.91\%$ & $2.82\%$ \\
    & Follow & $0.48\%$ & $0.33\%$ & $0.35\%$ \\
    & Forward & $0.21\%$ & $0.21\%$ & $0.28\%$ \\
    & Hate & $0.20\%$ & $0.06\%$ & $0.08\%$ \\
    & Click & $14.66\%$ & $58.38\%$ & $57.33\%$ \\
    & EffView & $45.57\%$ & $44.58\%$ & $48.48\%$ \\
    \midrule
    \multirow{3}{*}{\makecell[l]{User \\ Overlap}} 
    & Domain A & - & 63.64\% & 6.82\% \\
    & Domain B & 92.11\% & - & 9.09\%  \\
    & Domain C & 7.89\% & 7.27\%  & - \\
    \midrule
    \multirow{3}{*}{\makecell[l]{Item \\ Overlap}} 
    & Domain A & - & 38.54\% & 38.26\% \\
    & Domain B & 21.17\% & -  & 40.46\% \\
    & Domain C & 22.57\% & 43.43\% & - \\
    \bottomrule
    \end{tabular}}
    \vspace{-10px}
\end{table}

\begin{table*}[!]
\renewcommand\arraystretch{0.72}
\small
  \centering
  \caption{Performance comparison of different methods in terms of all six task metrics on three domains.
  The best and second-best results
  are highlighted in boldface and underlined respectively. $*$ indicates that the performance difference against the second-best result is statistically significant at $0.05$ level. The experimental results are averaged over five times.}
  \vspace{-5px}
  \label{tab::performanceA}
    \begin{tabular}{ccccccccccccc}
    \toprule
    \multirow{3}{*}{Method}&
    \multicolumn{6}{c}{Domain A | Double-Columned Discovery Tab (AUC)}&\multicolumn{6}{c}{Domain A | Double-Columned Discovery Tab (GAUC)}\cr
    \cmidrule(lr){2-7} \cmidrule(lr){8-13}
    &Like&Follow&Forward&Hate&Click&EffView
    &Like&Follow&Forward&Hate&Click&EffView \cr
    \midrule
    DeepFM & 0.8606 & 0.8025 & 0.7539  & 0.7092 & \textbf{0.6998} & 0.6908   & 0.6294 & 0.6401 & 0.6077 & 0.5490 & 0.5895  & 0.5815\cr
    DCN  & 0.8687 & 0.8017 & 0.7599  & 0.7178 & 0.6958 & 0.7038         & 0.6379 & 0.6533 & 0.6082 & 0.5378  & 0.5961 & 0.5893\cr
    xDeepFM  & 0.8706 & 0.8074 & \underline{0.7828}  & 0.7279  & 0.6961  & 0.7045   & 0.6459 &0.6525 & 0.6126 & 0.5319 & \textbf{0.5973}  & 0.5901 \cr
    DCNv2   & 0.8725  & \underline{0.8102} & 0.7615  & 0.7176 & 0.6973 & \underline{0.7046}  & 0.6441 & 0.6545 & 0.6161 & 0.5360 & \underline{0.5963} & \underline{0.5909} \cr
    \midrule
    DCNv2-MT   &0.8708 & 0.7949 & 0.7541  & 0.6489 & 0.6931 & 0.7007 & 0.6508 & 0.6468 & 0.6037 & 0.5187 & 0.5942 & 0.5907\cr
    SharedBottom  & 0.8685 & 0.7585 & 0.7587 & 0.7172 & 0.6922  & 0.7000  & 0.6301 & 0.6112 & 0.5782 & 0.4801 & 0.5933 & 0.5824\cr
    MMoE    & 0.8664 & 0.7676 & 0.7615     & 0.7306 & 0.6928 & 0.7010 &  0.6295 & 0.6155 & 0.5764 & 0.4998 & 0.5903 & 0.5806   \cr
    PLE   & \underline{0.8736} & 0.7991 & 0.7773     & \underline{0.7674}  & 0.6931  & 0.7006& 0.6337 & 0.6420 & 0.5854 &0.5338 & 0.5918  & 0.5812    \cr
    \midrule
    PLE-MD   & 0.8708 & 0.8001 & 0.7612  & 0.7310  & 0.6912  & 0.7041   & \textbf{0.6585} & 0.5985 & 0.5398  & 0.5455 & 0.5903 & 0.5881     \cr
    SharedTop & 0.8709 & 0.7973 & 0.7682 & 0.7601  & 0.6925 & 0.7035  & 0.6454& 0.6506 & \underline{0.6214} & \underline{0.5502} &  0.5936 &0.5872    \cr 
    SpecificTop   & 0.8700 & 0.7906 & 0.7624     & 0.7012  & 0.6928 & 0.7042 &0.6435 & \underline{0.6578} & 0.6131 & 0.4780 & 0.5939  & 0.5870    \cr
    SpecificAll    & 0.8673 & 0.7705 & 0.7618     & 0.7122 & 0.6926 & 0.7010  & 0.5924 & 0.6269 & 0.6076 & 0.5119 & 0.5621 & 0.5819     \cr
    \midrule
    PEPNet & \textbf{0.8797}* & \textbf{0.8258}* & \textbf{0.7911}* &  \textbf{0.7887}* & \underline{0.6957} & \textbf{0.7080}* & \underline{0.6549} & \textbf{0.6704}* & \textbf{0.6397}* & \textbf{0.5517} & 0.5950  & \textbf{0.5938}* \cr
    \bottomrule
    \end{tabular}
    \vspace{-10px}
\end{table*}
 
\begin{table*}[!]
\renewcommand\arraystretch{0.87}
\small
  \centering
  \label{tab::performanceB}
    \begin{tabular}{ccccccccccccc}
    \toprule
    \multirow{3}{*}{Method}&
    \multicolumn{6}{c}{Domain B | Featured-Video Tab (AUC)}&\multicolumn{6}{c}{Domain B | Featured-Video Tab (GAUC)}\cr
    \cmidrule(lr){2-7} \cmidrule(lr){8-13}
    &Like&Follow&Forward&Hate&Click&EffView
    &Like&Follow&Forward&Hate&Click&EffView \cr
    \midrule
    DeepFM   & 0.8901 & 0.8616 & 0.7738  & 0.8017 & 0.7156  & 0.7044   & 0.6247 &0.6388 &0.6020 & 0.5573 & 0.6106 & 0.6018\cr
    DCN  & 0.8949 & 0.8618 & 0.7783  & 0.8083  & 0.7152 & 0.7072   & 0.6342 & 0.6493 & 0.5992 & 0.5603 & 0.6105 & 0.6065 \cr
    xDeepFM   & 0.9027 & 0.8670  & \underline{0.7796}  & 0.8071 & \underline{0.7191} & \underline{0.7075}   & 0.6378 & \underline{0.6563} & 0.6006 & 0.5647 & 0.6109 & 0.6127\cr
    DCNv2   & \underline{0.9040} & 0.8601 & 0.7767  & 0.8111 & 0.7190 & 0.7072  & \underline{0.6408} & 0.6525 & 0.6059 & 0.5769 & \underline{0.6149}  & 0.6130\cr
    \midrule
    DCNv2-MT  & 0.9008 & 0.8523 & 0.7687  & 0.7886  & 0.7185  & 0.7074 & 0.6365 & 0.6465 & 0.6011 & 0.5716 & 0.6148  & \underline{0.6143} \cr
    SharedBottom   & 0.8876 & 0.8629 & 0.7746 &\underline{0.8399}  & 0.7154  & 0.7033  & 0.6267  & 0.6415 & 0.6060 & 0.5598 & 0.6098 & 0.6092\cr
    MMoE    & 0.8889 & 0.8611 & 0.7760   & 0.8325  & 0.7155 & 0.7037  & 0.6294 & 0.6499 & \underline{0.6061} & \underline{0.5841} & 0.6127 & 0.6126    \cr
    PLE   & 0.8905 & \underline{0.8677} & 0.7625     & 0.8326  & 0.7157& 0.7033 & 0.6304 & 0.6472 & 0.5939 & 0.5822  & 0.6106 & 0.6095     \cr
    \midrule
    PLE-MD   & 0.8606 & 0.7949 & 0.6184 & 0.7724 & 0.5288 & 0.5946  & 0.5712 & 0.6111 & 0.5251 & 0.5330 & 0.5666 & 0.5596    \cr
    SharedTop& 0.9002 & 0.8647 & 0.7705 & 0.8302  & 0.7185  & 0.7070  & 0.6239 & 0.6505 & 0.6001 & 0.5835 & 0.6125 & 0.6101    \cr 
    SpecificTop  & 0.8139 & 0.7534 & 0.6834     & 0.6525  & 0.3859 & 0.4016  & 0.5633 & 0.6033 & 0.5767 & 0.5224 & 0.4995 & 0.4996    \cr
    SpecificAll   & 0.8790  & 0.8565     & 0.7746 & 0.8300  & 0.7161 & 0.7044  & 0.6266 & 0.6411 & 0.6047 & 0.5640  & 0.6115  & 0.6119     \cr
    \midrule
    PEPNet & \textbf{0.9042} &\textbf{0.8837}* & \textbf{0.7974}* & \textbf{0.8587}* & \textbf{0.7203}* & \textbf{0.7092} & \textbf{0.6431} & \textbf{0.6705}* & \textbf{0.6257}* & \textbf{0.6207}* & \textbf{0.6189}* & \textbf{0.6208}* \cr
    \bottomrule
    \end{tabular}
    \vspace{-10px}
\end{table*}

\begin{table*}[!]
\renewcommand\arraystretch{0.87}
\small
  \centering
  \label{tab::performanceC}
    \begin{tabular}{ccccccccccccc}
    \toprule
    \multirow{3}{*}{Method}&
    \multicolumn{6}{c}{Domain C | Single-Columned Slide Tab (AUC)}&\multicolumn{6}{c}{Domain C | Single-Columned Slide Tab (GAUC)}\cr
    \cmidrule(lr){2-7} \cmidrule(lr){8-13}
    &Like&Follow&Forward&Hate&Click&EffView
    &Like&Follow&Forward&Hate&Click&EffView \cr
    \midrule
     DeepFM & 0.8945 & 0.8571 &0.7783  & 0.8406  & 0.7154   & 0.7107 & 0.6350 & 0.6379 & 0.6024  & 0.5763  &\textbf{0.6350} & 0.6202\cr
    DCN  & 0.8962 & 0.8598 & 0.7801  & 0.8431  & 0.7142  & 0.7136   & 0.6402 & 0.6451 & 0.6082 & 0.5805 & 0.6209 & 0.6231\cr
    xDeepFM   & 0.9013 & 0.8633 & 0.7796  & 0.8514 & 0.7192 & 0.7178 & 0.6431 & 0.6465 & 0.6055 & 0.5738 & 0.6227   & 0.6272\cr
    DCNv2  & \underline{0.9025} & 0.8603 & \underline{0.7806}  & \underline{0.8521} & 0.7261  & 0.7181 &     \underline{0.6455} & \underline{0.6505} & 0.6192 & 0.5827 & 0.6240  & 0.6292\cr
    \midrule
    DCNv2-MT   & 0.9022 & 0.8583 & 0.7710  & 0.8430 & \underline{0.7273}  & \underline{0.7182}  & 0.6442 & 0.6423 & 0.6093 & 0.5726 & 0.6247  & \underline{0.6296}\cr
    SharedBottom    & 0.9017 & 0.8574 & 0.7677  & 0.8346  & 0.7242  & 0.7152   & 0.6359 & 0.6436 & \underline{0.6194} & 0.5834 & 0.6222 & 0.6240\cr
    MMoE    & 0.9014 & 0.8565 & 0.7667     & 0.8432 & 0.7245 & 0.7143 & 0.6334 & 0.6370 & 0.6131 & 0.5663 & 0.6214 & 0.6232     \cr
    PLE   & 0.9018 & \underline{0.8651} & 0.7723     & 0.8507  & 0.7246 & 0.7155  & 0.6345 & 0.6467 & 0.6142 & \underline{0.6053} & 0.6233 & 0.6257     \cr
    \midrule
    PLE-MD    & 0.7237 & 0.7621 & 0.5203 &  0.7146 &  0.4437 & 0.4491 & 0.5432 & 0.5984 & 0.4770 & 0.4740 & 0.5470 & 0.5005     \cr
    SharedTop  & 0.9019 & 0.8605 & 0.7641  & 0.8458 & 0.7249 & 0.7180 & 0.6337 & 0.6424 & 0.6169 & 0.5863 & 0.6217  & 0.6271     \cr 
    SpecificTop    & 0.2056 & 0.6330 & 0.5199     & 0.6426 & 0.4833 & 0.4540  & 0.4214 & 0.5018 & 0.4778 & 0.5156 & 0.4919 & 0.4821   \cr
    SpecificAll    & 0.9011 & 0.8582 & 0.7683     &  0.8510 & 0.7244 & 0.7148  & 0.6333 & 0.6375 & 0.6174 & 0.5810 & 0.6208 & 0.6237  \cr
    \midrule
    PEPNet& \textbf{0.9063}* &\textbf{0.8843}* & \textbf{0.7927}* & \textbf{0.8589}* &\textbf{0.7296} & \textbf{0.7203} & \textbf{0.6501}* & \textbf{0.6720}* & \textbf{0.6373}* & \textbf{0.6212}* & \underline{0.6311}  & \textbf{0.6342}* \cr
    \bottomrule
    \end{tabular}
    \vspace{-12px}
\end{table*}

\vspace{-2px}
\subsubsection{Baselines and Implementations.} 
To demonstrate the effectiveness of PEPNet, we compare it with several state-of-the-art methods. 
The baselines fall into three categories: general recommenders that only deal with a single task on a single domain, multi-task recommenders that ignore the impact of multiple domains, and multi-task and multi-domain recommenders that consider comprehensive.

\noindent\textbf{General Recommenders:}
We train each task in each domain separately to report the multi-task and multi-domain results.
\begin{itemize}[leftmargin=*,partopsep=0pt,topsep=0pt]
\setlength{\itemsep}{0pt}
\setlength{\parsep}{0pt}
\setlength{\parskip}{0pt}
\item \textbf{DeepFM}
~\cite{guo2017deepfm} is a widely used general recommender, which replaced the wide part of WDL~\cite{cheng2016wide} with Factorization Machine.
\item \textbf{DCN}
~\cite{wang2017deep} replaces Factorization Machine of DeepFM with Cross Network to model the linear cross-feature.
\item \textbf{xDeepFM}
~\cite{lian2018xdeepfm} further introduces vector-wise idea into the Cross part of DCN to learn feature crosses efficiently.
\item \textbf{DCNv2} 
~\cite{wang2021dcn}
uses a mixture of low-rank DCN with a healthier trade-off between performance and latency to achieve SOTA.
\end{itemize}

\noindent\textbf{Multi-task Recommenders:}
We train multiple tasks in each domain separately to report the multi-task and multi-domain results.
\begin{itemize}[leftmargin=*,partopsep=0pt,topsep=0pt]
\setlength{\itemsep}{0pt}
\setlength{\parsep}{0pt}
\setlength{\parskip}{0pt}
	\item \textbf{DCNv2-MT}
 extends DCNv2 to multi-task scenarios, which shares the main model between different tasks and use different DNN layers to generate preference scores.
  \item \textbf{SharedBottom}
is the most common multi-task model that shares the parameters of the bottom DNN layers and uses specific task tower to generate corresponding scores.
\item \textbf{MMoE}
~\cite{ma2018modeling} shares several expert submodels and a gating network across all tasks to implicitly model relationships between multiple task with different label spaces.
\item \textbf{PLE}
~\cite{tang2020progressive}
is the state-of-the-art method which sets up independent experts for each task and considers the interaction between experts based on retaining the shared experts in MMoE.
\end{itemize}

\noindent\textbf{Multi-task and Multi-domain Recommenders:}
There is little work dedicated to solving multi-task and multi-domain recommendations at the same time.
We propose some variants to fill the gap.
\begin{itemize}[leftmargin=*,partopsep=0pt,topsep=0pt]
\setlength{\itemsep}{0pt}
\setlength{\parsep}{0pt}
\setlength{\parskip}{0pt}
	\item \textbf{PLE-MD} extends PLE to multi-domain scenarios, which shares the input Embedding layer across different domains.
  \item \textbf{SharedTop} first  shares input Embedding layer like PLE-MD, and shares the top DNN task towers across different domains, unlike SharedBottom which shares the bottom DNN layer.
	\item \textbf{SpecificTop}: 
	Different from SharedTop, this model adopts different task towers for the same task on different domains, while the bottom Embedding layer is still shared across domains.
	\item \textbf{SpecificAll}: 
	Different from SpecificTOP, this model not only distinguishes different top DNN task towers on different domains, but also adopts specific bottom Embedding layers.
\end{itemize}

\begin{figure*}[t]
\vspace{-0.3cm}
\centering
\hspace{-0.4cm}
\subfigure
[Like \& Follow Metric in Domain A, Forward \& Hate Metric in Domain B, and Click \& EffView Metric in Domain C
]{
\includegraphics[width=0.16\textwidth]{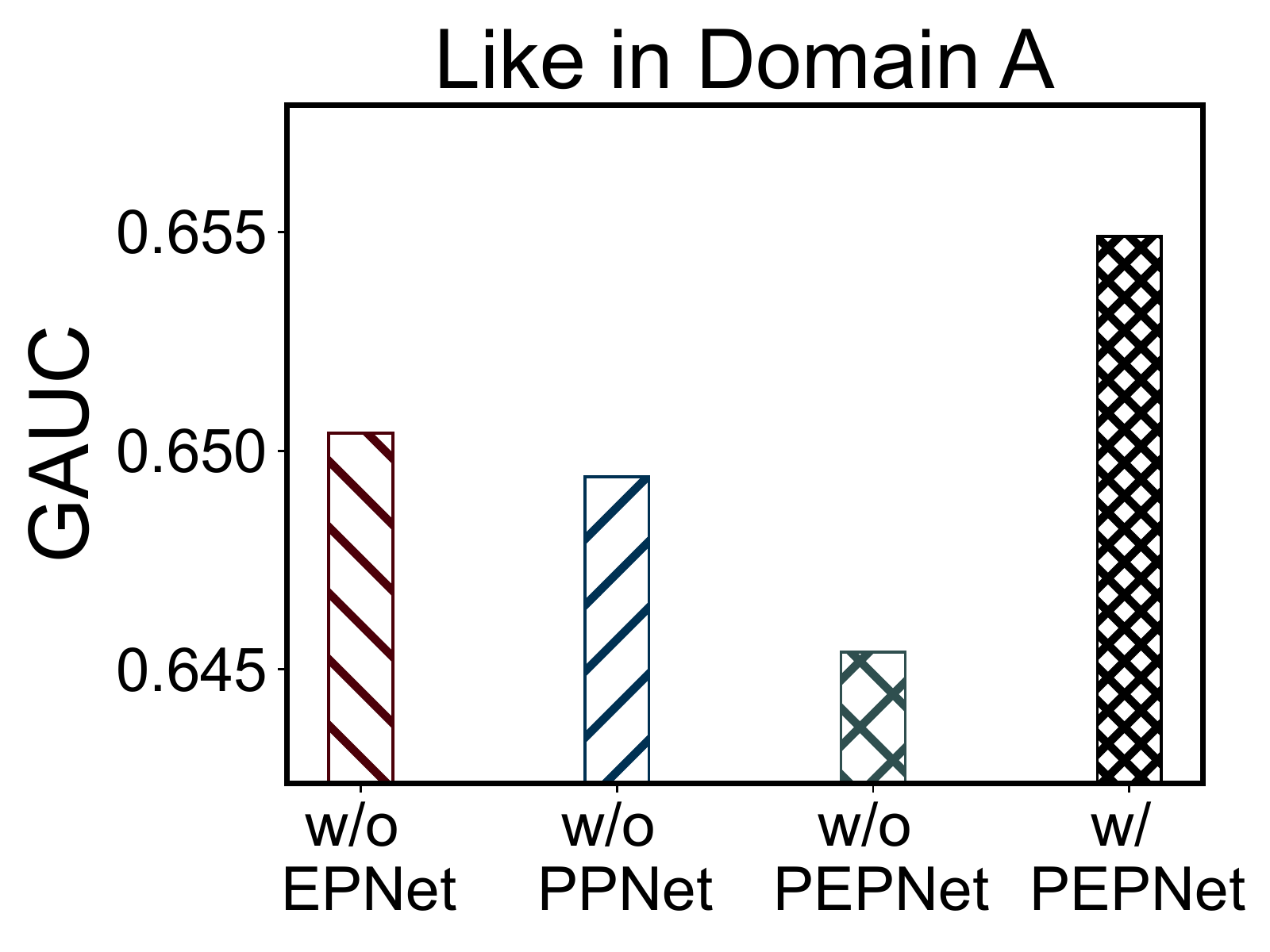}
\includegraphics[width=0.16\textwidth]{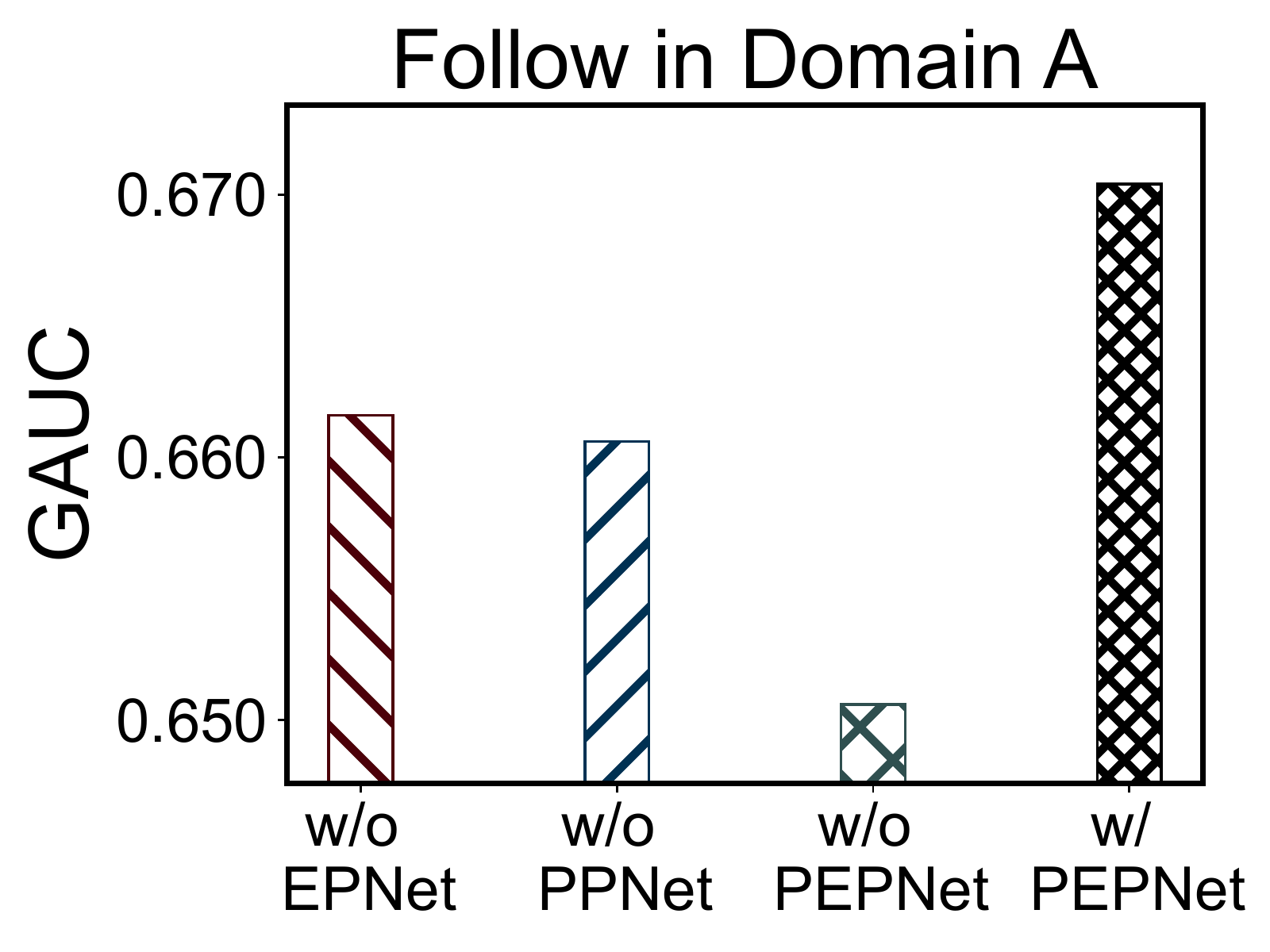}
\includegraphics[width=0.16\textwidth]{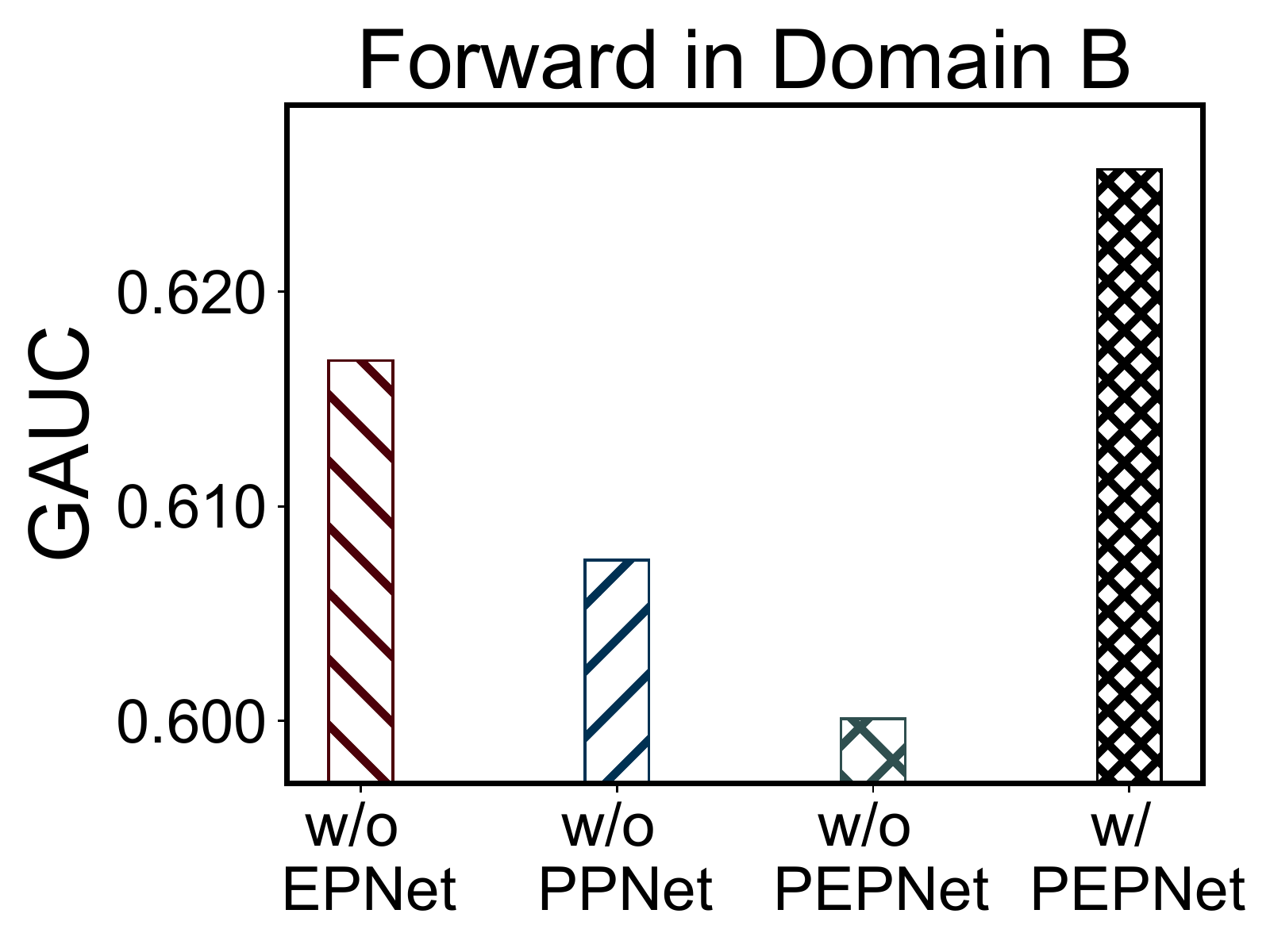}
\includegraphics[width=0.16\textwidth]{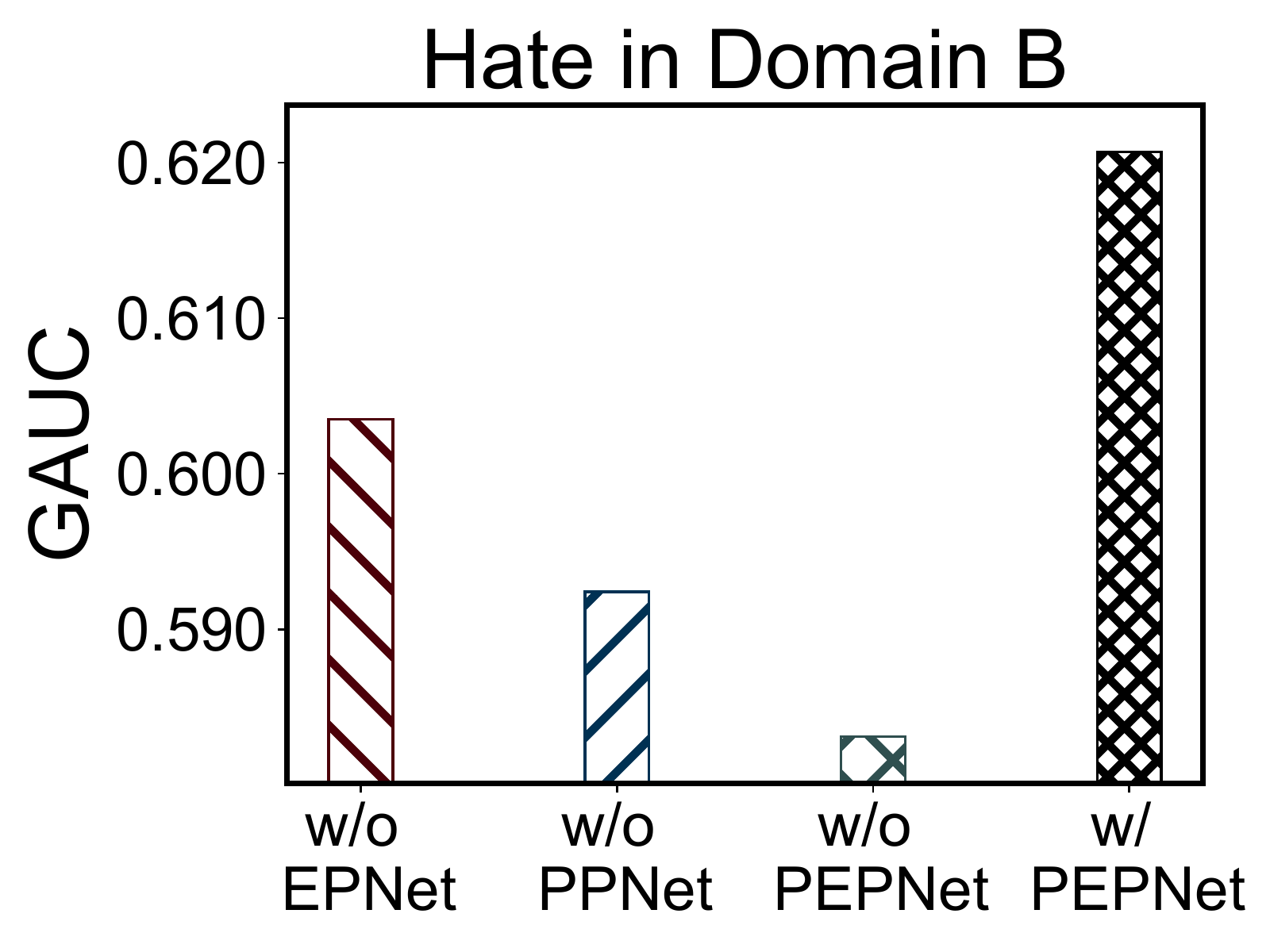}
\includegraphics[width=0.16\textwidth]{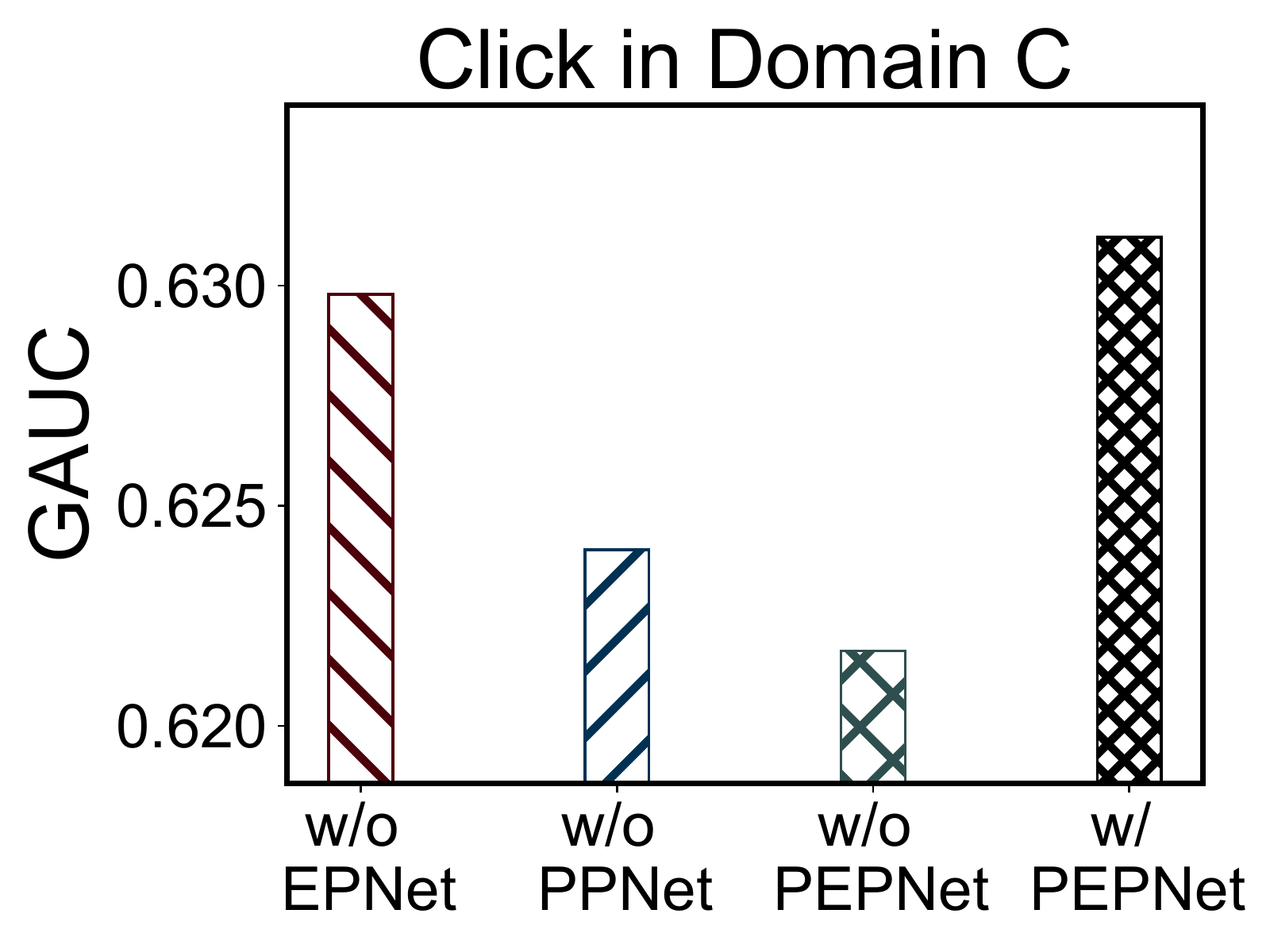}
\includegraphics[width=0.16\textwidth]{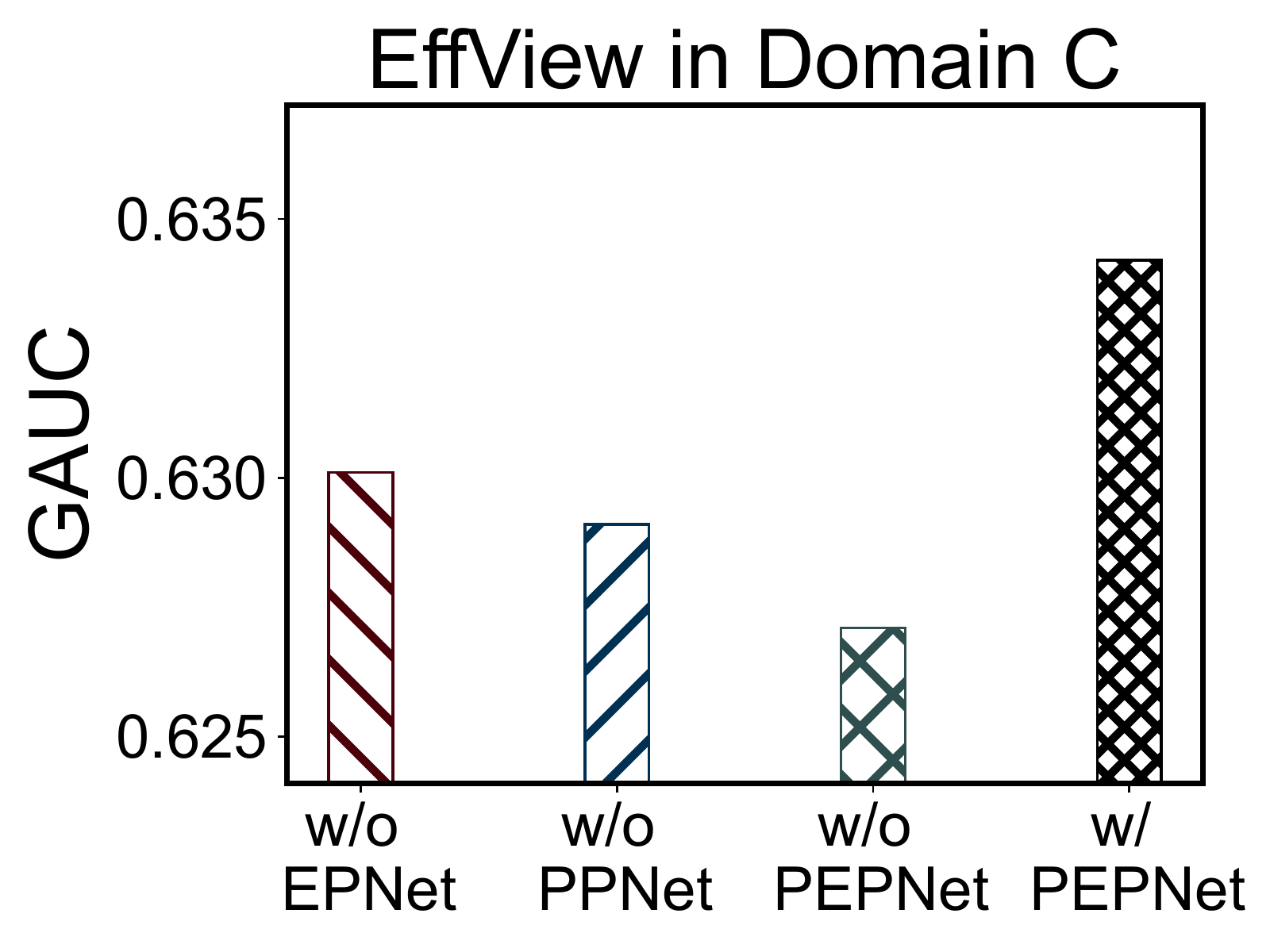}
} 
\\
\vspace{-0.1cm}
\subfigure
[All task metrics in Domain C]{
\includegraphics[height=0.11\textwidth]{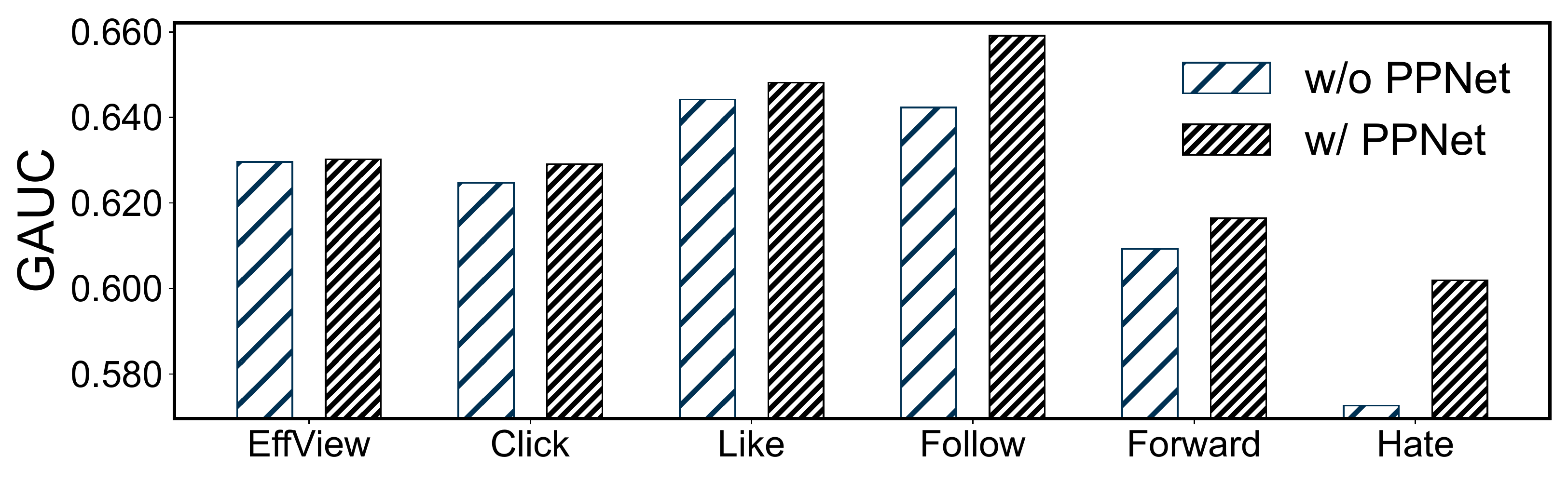}
} \label{fig::ablation1}
\subfigure
[Follow Metric in all domains]{
\includegraphics[height=0.11\textwidth]{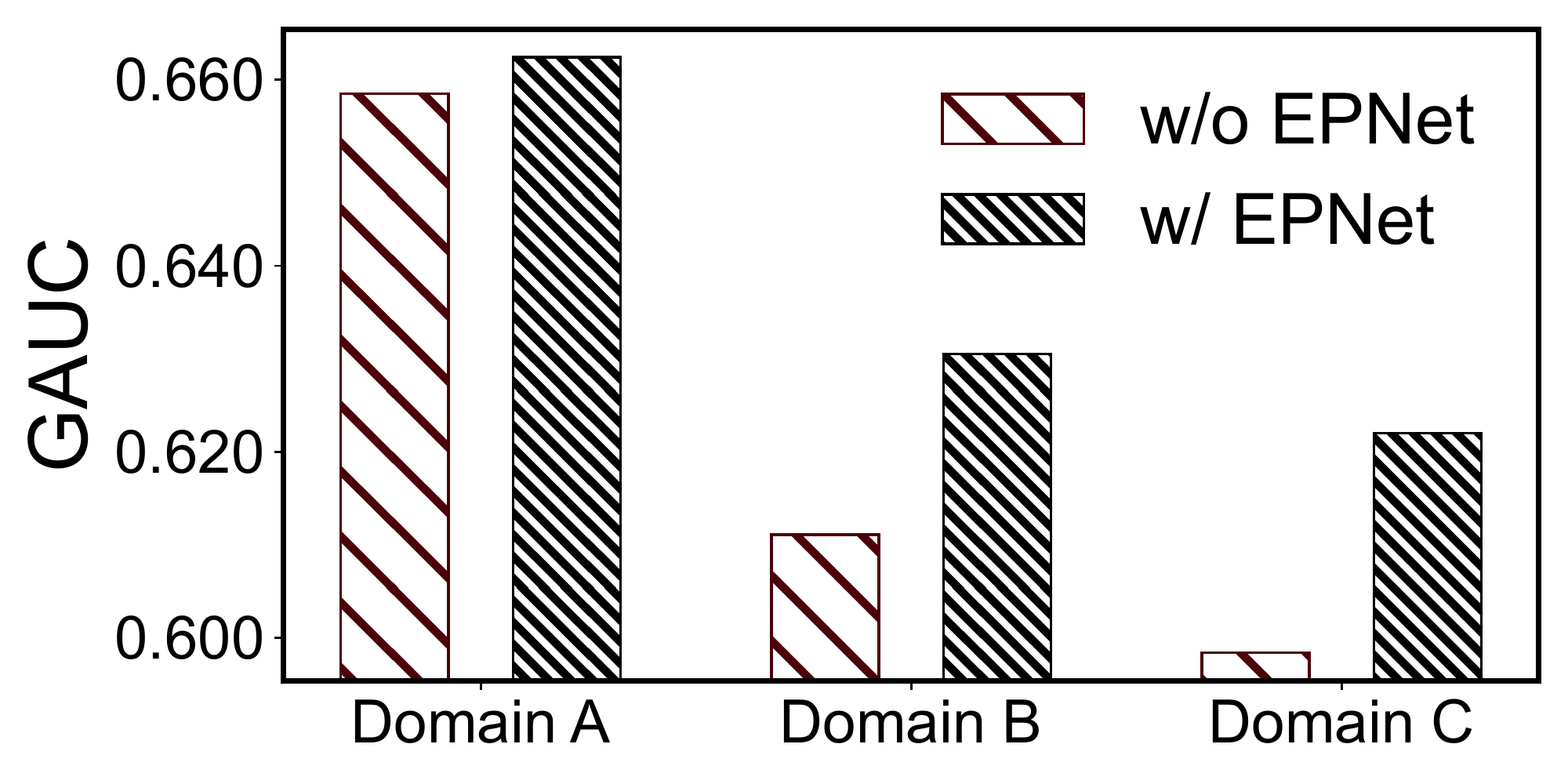}
}
\subfigure
[Hate Metric in Domain C]{
\includegraphics[height=0.11\textwidth]{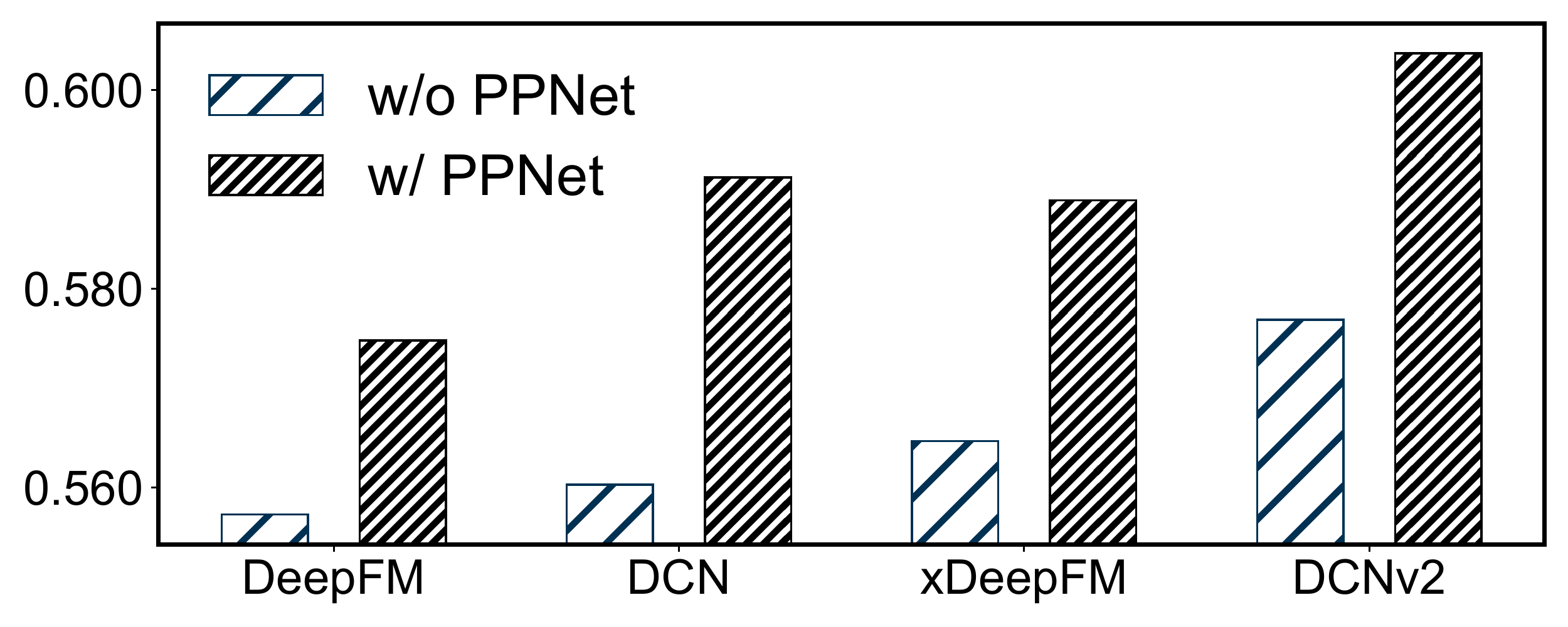}
}
\vspace{-0.4cm}
\caption{Effectiveness of the sub-modules PPNet and EPNet in the proposed PEPNet model. 
}
\vspace{-0.4cm}
\label{fig::ablation}
\end{figure*}

\begin{figure*}[t]
\centering
\subfigure
[Embedding dims in EPNet]{
\includegraphics[width=0.21\textwidth]{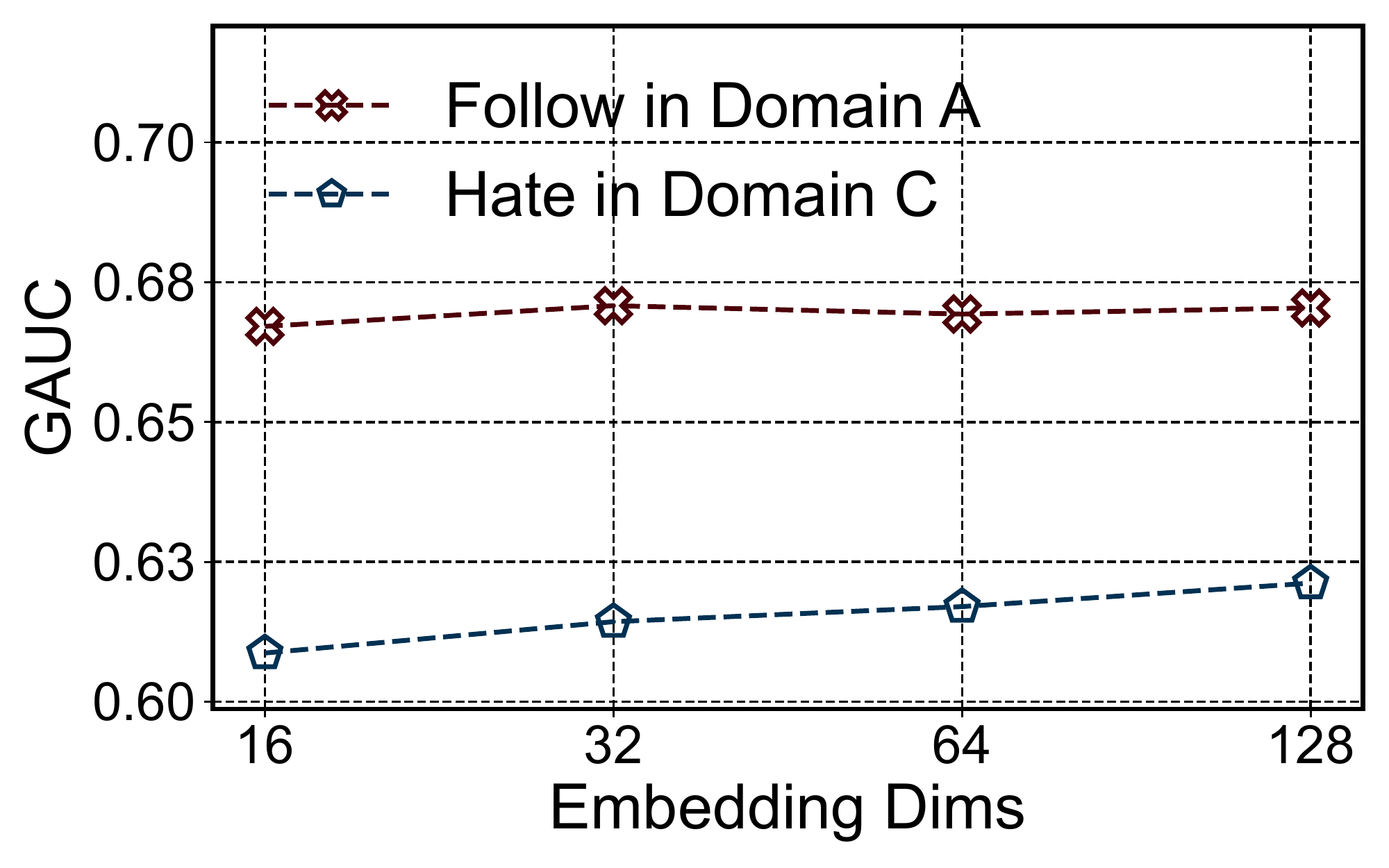}
}
\subfigure
[DNN layers in PPNet]{
\includegraphics[width=0.21\textwidth]{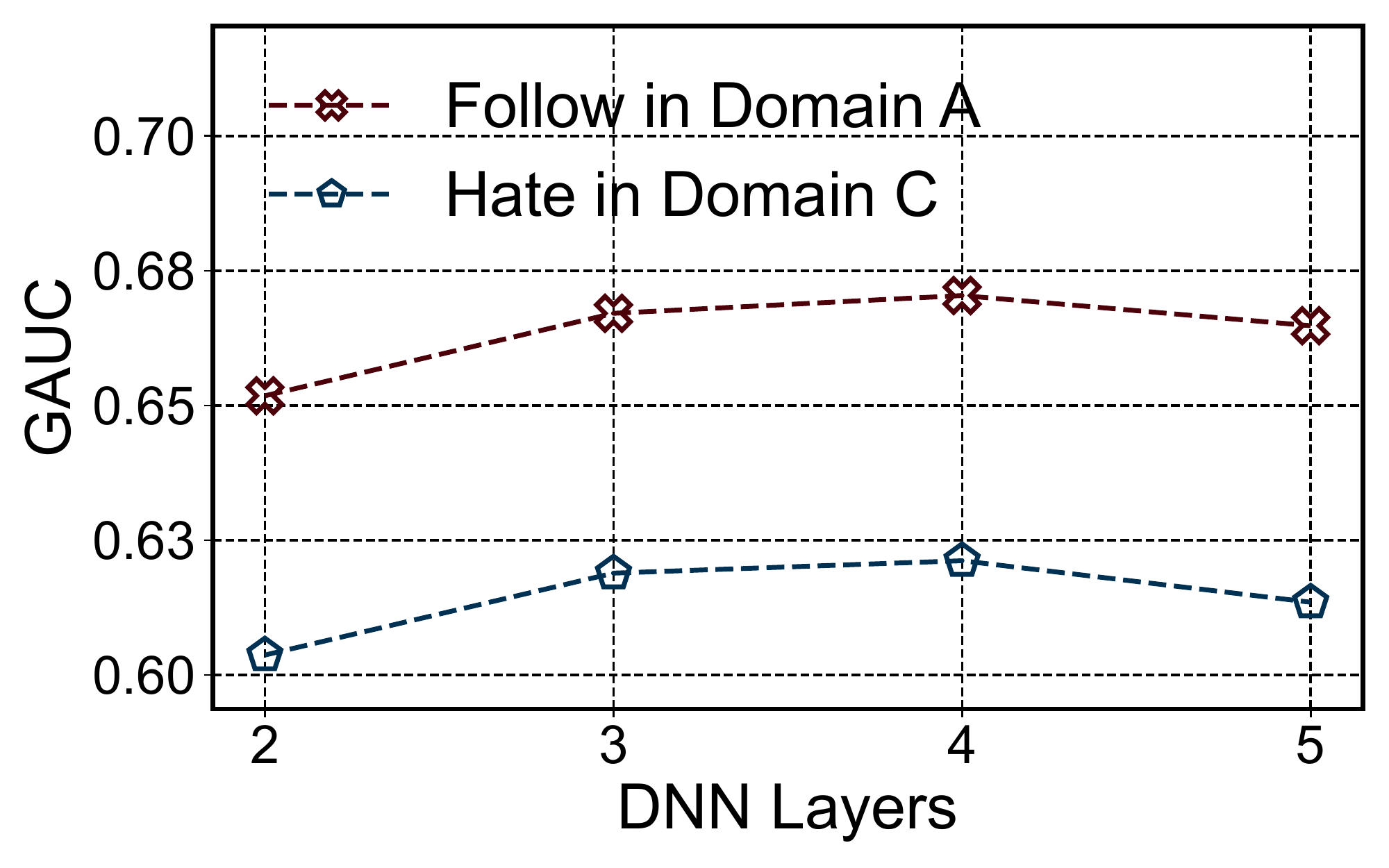}
}
\subfigure
[Coefficients in Gate NU]{
\includegraphics[width=0.21\textwidth]{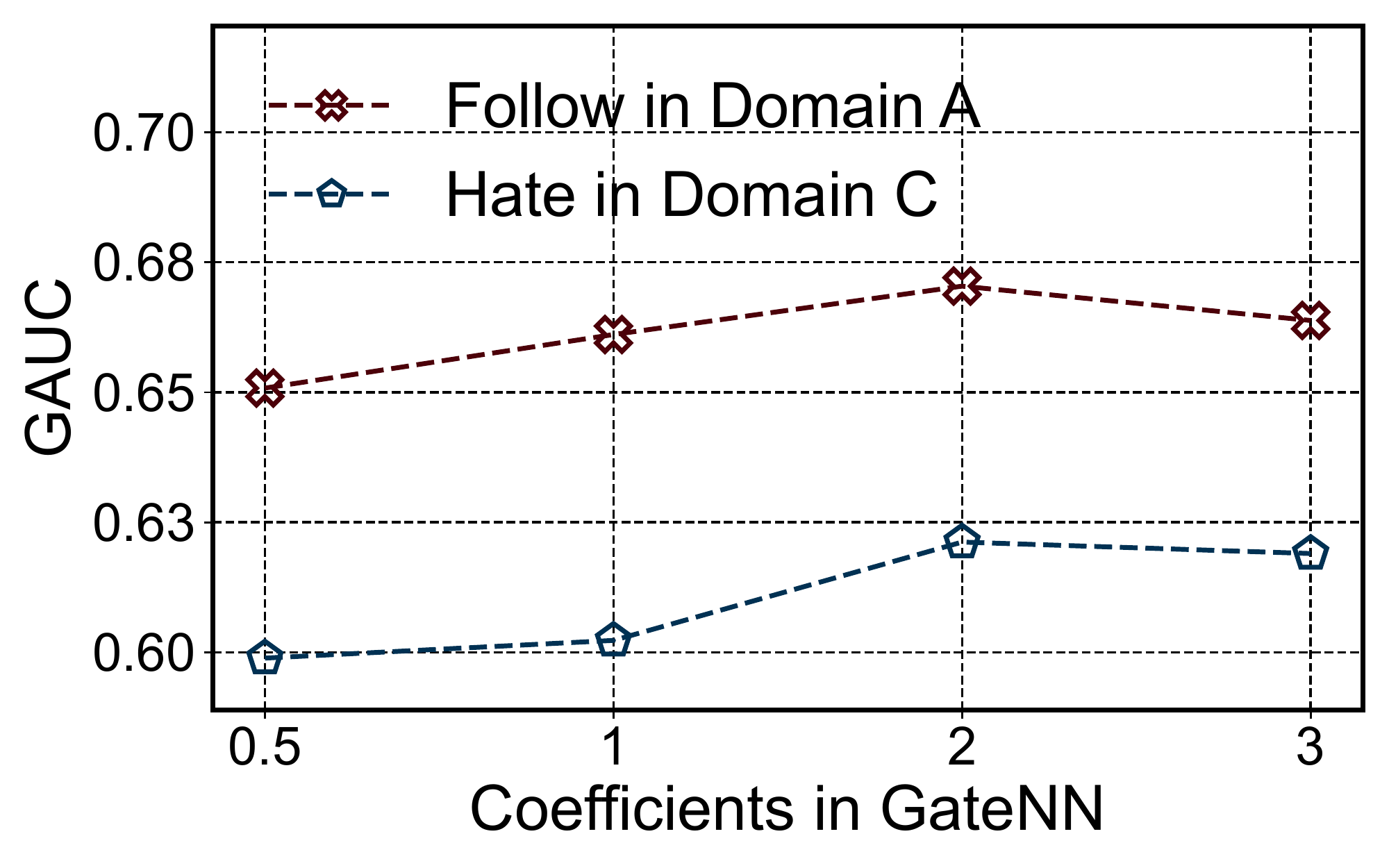}
}
\subfigure
[Extra Input and BP]{
\includegraphics[width=0.21\textwidth]{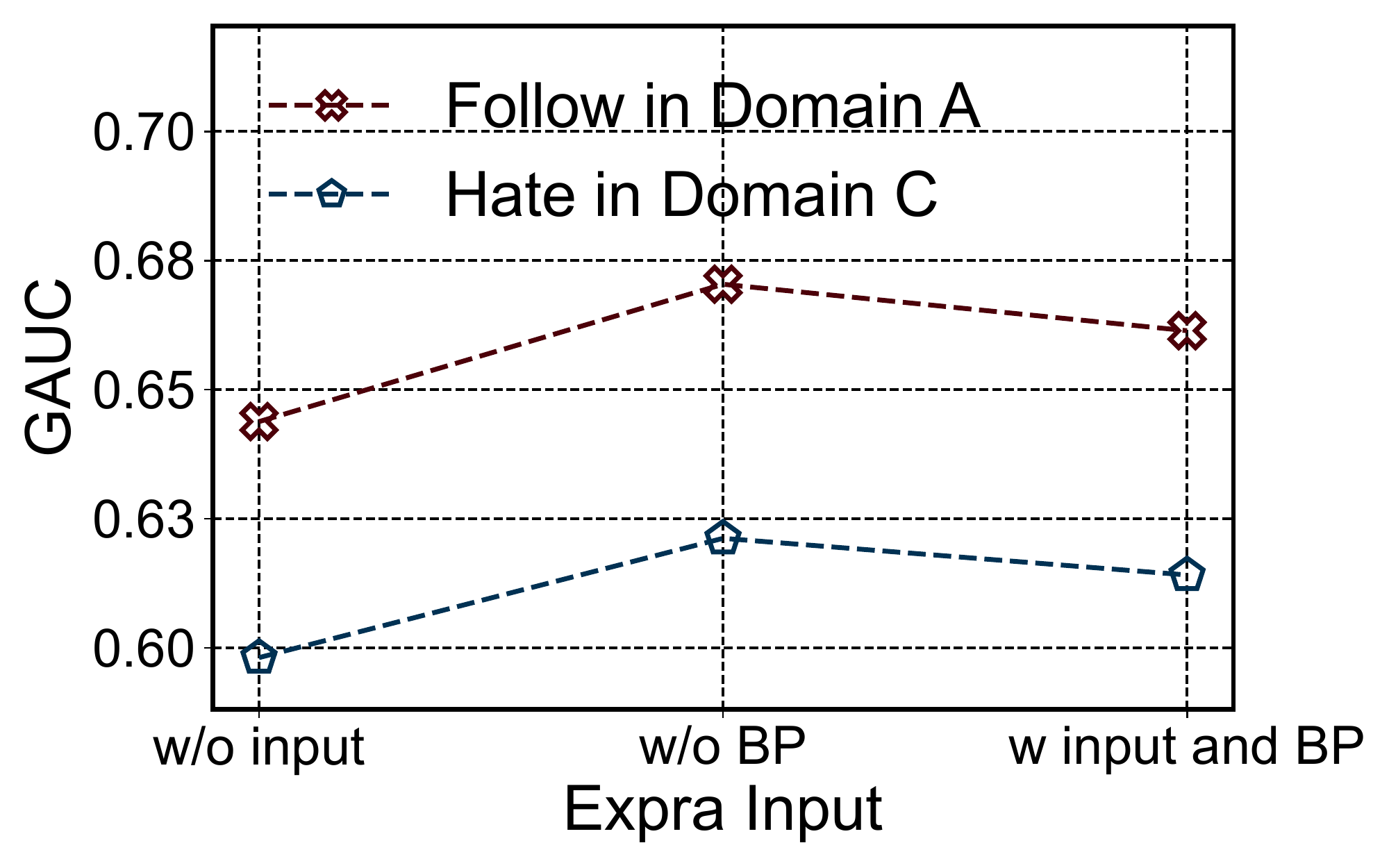}
}
\\
\vspace{-0.4cm}
\caption{Performance of PEPNet model with different settings and implementations.
}
\vspace{-0.4cm}
\label{fig::hyperpara}
\end{figure*}

\vspace{-8px}
\subsubsection{Hyper-parameter Settings.} 
In offline experiments, we implement all the models based on TensorFlow\cite{abadi2016tensorflow}. 
We use Adam~\cite{Adam} for optimization with the initial learning rate as 0.001. 
The batch size is set as 1024 and the embedding size is fixed to 40 for all models. 
Xavier initialization~\cite{xavier} is used here to initialize the parameters. 
All methods use a two-layer feedforward neural network with hidden sizes of [100, 64] for interaction estimation. 
The prior information used in EPNet and PPNet is added as additional inputs to the Embedding layer in all baselines for a fair comparison.
We apply careful grid-search to find the best hyper-parameters. 
The number of experts in MMoE, PLE and its variants is searched in [4, 6, 8].
All regularization coefficients are searched in $[1e^{-7}, 1e^{-5}, 1e^{-3}]$.

\vspace{-10px}
\subsection{Overall Performance (RQ1)}
Table~\ref{tab::performanceA}
illustrates the experimental results of six tasks in three domains.
From the results, we have the following observations:

\begin{itemize}[leftmargin=*,partopsep=0pt,topsep=0pt]
\setlength{\itemsep}{0pt}
\setlength{\parsep}{0pt}
\setlength{\parskip}{0pt}
\item \textbf{Our proposed method consistently achieves the best performance.}
We can observe that our model PEPNet significantly outperforms all baselines in terms of all six task metrics on three domains.
Specifically, our model improves GAUC on average by around 0.01 on Domain A, 0.02 on Domain B and 0.02 on Domain C with \textit{p}-value < 0.05.
For the average performance of each task on three domains, Like is increased by 0.01, Follow is increased by 0.02, Forward is increased by 0.02, Hate is increased by 0.03, Click is increased by 0.002, and EffView is increased by 0.005.
The improvement is more obvious on the more sparse domain and task, which verifies that our method can balance multi-task and multi-domain recommendation problems more effectively. 
It significantly reduces the difficulty of modeling sparse domains and sparse tasks in a cross-domain and cross-task manner.

\item \textbf{General recommenders cannot balance the task seesaw.}
The general recommender performs well on dense tasks (Click) in dense domains (Domain B), but performs poorly on sparse tasks (Forward) in sparse domains (Domain A).
Simply extending the general recommender (DCNv2) to multi-task (DCNv2-ML) results in some tasks (Like) getting better and some tasks (Hate) getting worse.
This shows that the centralized general model has seesaw problems when faced with multi-task estimation, resulting in unbalanced performance across tasks.
Compared with this, SharedBottom with a shared parameter layer and specific task towers obtains balanced performance improvements in all metrics on some domains (Domain C).
It demonstrates that specially designed multi-task recommenders can alleviate the task seesaw phenomenon.
And the more complex the design of the shared parts and specific parts of the model (MMoE and PLE), the more obvious the performance improvement.
But they still suffer from poor performance on sparse domains (Domain A).

\item \textbf{Multi-task recommenders cannot balance the domain seesaw.}
Even the most powerful multi-task recommenders (PLE), when extended to multi-domain (PLE-MD),
still appears that some domains (Domain A) get better and some domains (Domain C) get worse, i.e. the domain seesaw phenomenon.
The reason is that the top-level label space and the bottom-level embedding space have inconsistencies.
The limitation of multi-task methods that model separate domains is that they cannot consider cross-domain and cross-task information simultaneously.
The multi-task and multi-domain variants SharedTop built on SharedBottom, an early effort in multi-task learning, can alleviate the double seesaw phenomenon to a certain extent.
With the task tower being specific to domains, SpecificTop only brings better results in some domains (Domain A) while increasing the number of parameters several times.
While SpecificAll further divides the underlying embedding space, which ignores the shared knowledge between domains and deteriorates the recommendation effect.
Our method plugs gated networks based on shared bottom Embedding layers and shared top DNN task towers to capture the user's personalized bias across domains and tasks, achieving the best performance with a small number of parameters.
\end{itemize}

\vspace{-10px}
\subsection{Ablation Study (RQ2)}
To further validate the effectiveness of the sub-modules proposed in the PEPNet model, we compare the offline performance of models without the PPNet module, without the EPNet module, without both modules, and the complete model, as shown in Figure~\ref{fig::ablation} (a).
In addition, we study the generalization ability of PEPNet as a plug-and-play module on other settings than multi-task and multi-domain recommendation problems.
Specifically, we compare the effect of PPNet for multi-task and single-domain recommendation in Figure~\ref{fig::ablation} (b), the effect of EPNet for single-task and multi-domain recommendation in Figure~\ref{fig::ablation} (c), and the effect of adding PPNet to single-task and single-domain models in Figure~\ref{fig::ablation} (d).

The results in Figure~\ref{fig::ablation} (a), (b) and (c)  show the effectiveness of capturing cross-domain and cross-task information via EPNet and PPNet.
Embedding personalization of EPNet and parameter personalization of PPNet can bring further performance improvement, respectively.
In Figure~\ref{fig::ablation} (d), adding pure parameter personalization to single-task and single-domain models can also bring benefits for general recommendation problems, which also illustrates the importance of modeling personalization bias in recommendation.

\vspace{-7px}
\subsection{Hyper-parameter Study (RQ3)}
To study the influence of different settings and implementations in the proposed model, we conduct hyperparameter experiments.
Firstly, we compare the performance of EPNet under different embedding sizes for each input feature in Figure~\ref{fig::hyperpara} (a), and the effect of the number of DNN layers coupled with PPNET in Figure~\ref{fig::hyperpara} (b).
Secondly, since we propose to add scaling factors on the Sigmoid in Gate NU to amplify or compress the differences between dimensions, we evaluate the recommendation performance under different coefficients in Figure~\ref{fig::hyperpara} (c).
Finally, we study the role of extra input in EPNet and PPNet, and compare the effect on performance of removing input, adding input but removing BackPropagation(BP), and adding input and BP in Figure~\ref{fig::hyperpara} (d).

From the results, we can observe that the performance of EPNet is robust under embeddings of different dimensions, and even small dimension with only 16 still keeps an excellent performance.
As the number of DNN layers increases, the performance of PPNet becomes better, but after a certain number of layers, a too deep neural network will lead to overfitting.
The coefficient of Sigmoid in Gate NU performs best when the value is $2$, because its output range is $(0,2)$ centered at $1$, which can better balance the scaling effect.
Adding general input and removing BackPropagation(BP) in EPNet and PPNet is better than other settings, which shows that this manner can make better use of input information and model user personalization without affecting the backbone network.

\vspace{-9px}
\subsection{Online A/B Testing (RQ4)}
To evaluate the online performance of PEPNet, we conduct rigorous online A/B testing. 
Table \ref{online} shows the improvements of three representative domains: Double-Columned Discovery Tab, Featured-Video Tab, and Single-Columned Slide Tab. 
Unlike CTR and GMV in e-commerce scenarios, short-video scenarios pay attention to the following metrics: Like, Follow, Forward and Watch Time. 
Watch Time measures the average watching time on videos of each user.
We can see that all metrics improve significantly compared with the previous SOTA method.
Note that 0.1\% increase in Watch Time is considered an effective improvement in Kuaishou, so PEPNet achieves significant business benefits.
PEPNet is deployed in our online service, serving over 300 million users every day.  

\vspace{-4px}
\section{Relate Work}
Our work builds on traditional CTR prediction and extends it to multi-domain and multi-task using gates.
In this section, we discuss related works about CTR prediction, multi-domain learning, multi-task learning, and gating mechanisms in the recommendation.

\vspace{-6px}
\subsection{\textbf{Click-Through Rate Prediction}}
Click-Through Rate(CTR) Prediction is the most important growth engine for E-commerce and streaming Internet companies, which can improve user experience and increase company revenue. The traditional shallow CTR models, e.g. Logistic Regression (LR), Factorization Machine (FM) and Gradient Boosting Decision Tree (GBDT), with their strong interpretability and lightweight training deployment requirements, were widely used in the early days.

Due to the powerful ability of deep learning to capture high-order feature cross, 
modern deep methods achieve significant improvements.
FNN\cite{zhang2016deep} uses FM to pre-train the embedding layer and then inputs the processed dense features into DNN.
PNN~\cite{qu2016product} transfers the vector inner/outer product from pre-training directly to the neural network.
WDL~\cite{cheng2016wide} jointly trains a wide linear model and a deep neural network to combine the memory and generalization advantages.
DeepFM~\cite{guo2017deepfm} replaces the wide part of WDL with FM, thus no longer relying on manual feature engineering.
DCN\cite{wang2017deep, wang2021dcn} replaces FM of DeepFM with Cross Network, and xDeepFM\cite{lian2018xdeepfm} further introduces the idea of vector-wise into the Cross part of DCN.
DCNv2~\cite{wang2021dcn}
uses a mixture of low-rank DCN with a healthier trade-off between performance and latency to achieve SOTA.

\begin{table}[]
\small
\caption{Online gains in three representative domain. Note that in the Kuaishou short-video recommendation scenario, 0.1\% increase in Watch Time is a significant improvement.}
\vspace{-5px}
\begin{tabular}{lccc}
\toprule
& Discovery Tab & Featured-Video Tab & Slide Tab \\
\midrule
Like   & +1.08\% & +1.36\% & +2.11\%  \\
Follow   & +1.43\% & +1.81\% & +2.23\%  \\
Forward   & +1.31\% & +1.55\% & +1.43\% \\
Watch Time   & +1.25\% & +1.93\% & +2.12\% \\
\bottomrule
\label{online}
\end{tabular}
\vspace{-20px}
\end{table}

\subsection{\textbf{Multi-domian Learning}}
Multi-domain learning is an extension of domain adaptation and belongs to transductive transfer learning. Transfer learning can use source domains with sufficient labeled data to help target domains with little labeled data.
When the data distributions in the source and target domains are different, but the two tasks are the same, this special kind of transfer learning is named Domain Adaptation~\cite{daume2006domain}.
Models trained directly on the source domain generally perform poorly on the target domain due to not satisfying the Independent and Identically Distributed (IID) assumption, a phenomenon known as Negative Transfer\cite{
ben2007analysis, ben2010theory}. 
The basic idea of domain adaptation is to align the data of different distributions of source and target domains into a unified space to obtain domain-invariant features.
Different from general domain adaptation problems, multi-source domain adaptation involves multiple source domains with different distributions, and multi-target domain adaptation aims to transfer to multiple target domains~\cite{zhu2019aligning}.
And the key to solving such problems is alignment strategies on multiple domains~\cite{zhao2020multi, peng2019moment, zhao2019multi}.

Traditional CTR prediction mainly focuses on estimating a single target in a single domain. With the continuous increase of real-world scenarios, joint training from different domain data needs to be considered.
Therefore, unlike previous work, multi-domain learning in recommendation scenarios \cite{sheng2021one, li2020ddtcdr} weakens the concepts of source and target domain and emphasizes the simultaneous improvement of recommendation effect in multiple domains.

\subsection{\textbf{Multi-task Learning}}
Multi-task learning aims to learn multiple related tasks at the same time, and facilitate the learning of each specific task by mining shared information. Early linear models~\cite{ArgyriouEP2008ConvexMultitask} used shared sparse representations to learn across multiple tasks. 
In the era of deep learning, hard parameter sharing methods may cause negative transfer due to task differences. To achieve better performance, some studies deal with optimization with soft parameter sharing methods. 
The cross-stitch network~\cite{MisraSGH2016CrossStitch} and sluice network~\cite{ruder2017sluice} are proposed to learn linear combinations of task-specific hidden layers.
Other methods use gating mechanisms and attention mechanisms for information fusion. MOE~\cite{jacobs1991adaptive} uses the gate structure to combine several experts shared at the bottom. MTAN~\cite{liu2019end} consists of a shared network and several task-specific attention modules.

In recommendation systems, early models based on collaborative filtering and matrix factorization \cite{wang2013online, wang2018explainable, lu2018like} 
express lower expressivity and ignore the correlation between tasks.
Due to irreplaceable advantages such as simplicity and efficiency, the hard parameter sharing at the bottom (ShareBottom) is widely used in recommendation systems.
MMoE\cite{ma2018modeling} further shares all experts in different tasks and use different gates for each task to extend MOE.
ESSM~\cite{ma2018entire} is based on a soft parameter sharing structure and simultaneously optimizes two related tasks with sequential modes to alleviate the sparsity of the prediction target.
Based on retaining the shared experts in MMoE, PLE~\cite{tang2020progressive} sets up independent experts for each task and considers the interaction between different experts.

\subsection{\textbf{Gating Mechanisms in Recommendation}}
Gating mechanism is widely used in recommendation systems due to its ability to adaptively strengthen important information and weaken irrelevant information. 
Most recently, \citet{ma2019hierarchical} propose a hierarchical gating network (HGN), which uses feature-level gate and instance-level gate modules to automatically model user's selection of item instances with different features. 
\citet{huang2019fibinet} applies the squeeze-excitation network (SENET) proposed in the field of computer vision to dynamically capture the importance of features and use a bilinear function to learn feature combinations. 
\citet{huang2020gatenet} propose the feature embedding gate and hidden gate, which adaptively select features and feature interactions to be passed to deeper layers of the network by using themselves as gate inputs. 
However, these methods emphasize information selection without considering personalized modeling.
Due to the lack of insight and design of domain seesaw and task seesaw, they are not suitable for the multi-task and multi-domain recommendation.

\section{Conclusion}
In this paper, we studied the imperfectly double seesaw problem, where some domains have much less data than others and some tasks suffer from sparse labels. Then, we proposed a Parameter and Embedding Personalized Network (PEPNet) which learned the heterogeneous relationships between multiple domains and multiple tasks. In Kuaishou's recommendation scene, embedding personalization and parameter personalization were fully considered, which greatly improved the user's consumption experience. And for the characteristics of short-video recommendation, we made engineering strategies to optimize during training and online inferring. We have deployed the model at Kuaishou Apps. All online and offline experiments of multiple tasks from multiple domains achieved significant improvements in both App usage and engagement.

\clearpage
\bibliographystyle{ACM-Reference-Format}
\balance
\bibliography{sample-base}

\end{document}